\documentclass[pra,twocolumn,showpacs,floatfix]{revtex4}
\usepackage{graphicx}
\usepackage{color}
\usepackage{amsmath}
\begin{document}

\title{Transition from the mean-field to the bosonic Laughlin state in a rotating Bose-Einstein condensate}

\author{G. Vasilakis$^1$, A. Roussou$^2$, J. Smyrnakis$^1$, M. Magiropoulos$^1$, W. von Klitzing$^3$, and G. M. Kavoulakis$^1$}
\affiliation{$^1$Hellenic Mediterranean University, P.O. Box 1939, GR-71004, 
Heraklion, Greece \\
$^2$Department of Mathematics and Applied Mathematics, University of Crete, GR-71004, Heraklion, Greece \\
$^3$Institute of Electronic Structure and Laser (IESL), Foundation for Research and 
Technology (FORTH), N. Plastira 100, Vassilika Vouton, 70013, Heraklion, Crete, Greece}

\begin{abstract}

We consider a weakly-interacting Bose-Einstein condensate that rotates in either a harmonic, or a 
weakly-anharmonic trapping potential. Performing numerical calculations, we investigate the behaviour of 
the gas in these two cases as the angular momentum, or equivalently as the rotational frequency of the trap 
increases. While in the case of a purely-harmonic potential the gas makes a transition from the mean-field 
regime to the correlated, ``Laughlin", regime, in the case of anharmonic confinement the mean-field 
approximation remains always valid. We compare our derived results in these two cases, using both the 
mean-field approximation, as well as the diagonalization of the many-body Hamiltonian considering a small 
atom number. 

\end{abstract}

\pacs{03.75.Lm, 05.30.Jp, 67.85.Hj, 67.85.Jk}

\maketitle

\section{Introduction}

One of the main characteristics of cold-atomic systems is that, under typical conditions, they are dilute, i.e., 
the scattering length for atom-atom elastic collisions is much smaller than the interparticle spacing. Under these
conditions the system is well described by the mean-field approximation. In the case of bosonic atoms, within this 
approximation the many-body state is simply a product state, and this leads to the well-known Gross-Pitaevskii 
equation, which is a nonlinear Schr\"odinger equation. 

The experimental realization of correlated states, where the mean-field approximation breaks down, as well as
their theoretical study, have attracted a lot of attention in recent years. Candidate systems for the realization 
of such states include Bose-Einstein condensates in harmonic and anharmonic traps under rotation, see e.g., 
Refs.\,\cite{qh1, qh1p, qh2, qh22, qh3, qh4, qh4p, qh4pp, qh5, qh6, qh7, qh8, qh9, qh10, qh11, qh12, Dalibard, 
Cornell} and \cite{qhanh1, qhanh2}, in optical lattices, see, e.g., Refs.\,\cite{optlat}, and in artificial 
gauge potentials see, e.g., Refs.\,\cite{artf}, etc.

In the case of rotation in a harmonic trapping potential there is an interesting transition as the angular 
frequency of the trap, or equivalently as the angular momentum of the gas increases. One may distinguish between 
two regimes, the limit of ``slow" and ``fast" rotation. In the first regime the angular frequency of the trap 
$\Omega$ is smaller than the trap frequency $\omega$, while the total angular momentum of the system $L$ is of 
order the total number of atoms $N$. In the second regime $\Omega$ approaches $\omega$, while $L$ becomes of order 
$N^2$. 

In the limit of slow rotation the system is well described by the mean-field approximation. The correlations
beyond the mean-field approximation lower the energy of the system to subleading order in $N$. On the
other hand, in the limit of rapid rotation, the correlations play a crucial role. In this case even the 
leading-order term of the energy is affected and, remarkably, the interaction energy reaches its absolute 
minimum, i.e., it vanishes, for $L \ge N(N-1)$. Actually, the many-body state of the system for $L = N(N-1)$ 
is given analytically by the bosonic version of the Laughlin state \cite{LS} that appears in the quantum Hall 
effect \cite{HE, HErev}. We stress that in this regime the mean-field approximation cannot in any way describe 
accurately the lowest-energy state of the system.

In the case of rotation in an anharmonic trapping potential, which we also examine in the present study, for
slow rotational frequencies of the trap the system is well described by the mean-field approximation. Depending 
on the value of the coupling and the angular velocity, the lowest-energy state consists either of giant vortices, 
a vortex lattice, or a combination of the two, i.e., a vortex lattice with a ``hole" at the center of the trap 
\cite{SF, EL, U, FB, KB, JK, Str, fan, Zar, Corr}. As we have shown recently \cite{frag}, in the limit of fast 
rotation the Laughlin-like states which one gets in the case of a purely harmonic potential for $L \ge N(N-1)$ 
are highly fragile against giant-vortex states, in the presence of even a very ``weakly"-anharmonic potential.

In the present study we consider a Bose-Einstein condensate that rotates either in a harmonic trapping potential,
or in a weakly-anharmonic potential and study the transition from the limit of slow to the limit of fast rotation. 
More specifically, in what follows below we describe in Sec.\,II the two basic approaches that we use, namely the 
mean-field approximation, as well as the method of diagonalization of the many-body Hamiltonian.  

Then, in Sec.\,III, we evaluate the energy and the rotational response of this system using both the diagonalization,
as well as the mean-field approximation for the cases of harmonic and anharmonic confinement. We proceed in Sec.\,IV
with the evaluation of the single-particle density distribution and the pair-correlation function that results from
the diagonalization in both a harmonic and in an anharmonic trapping potential. In Sec.\,V we present the results
from the overlap of the many-body state in the presence and in the absence of the anharmonic potential, as well as
the eigenvalues of the density matrix in the two cases. Finally in Sec.\,VI we describe a summary of our study and 
our conclusions.

\begin{figure}[t]
\includegraphics[width=7.5cm,height=5cm,angle=0]{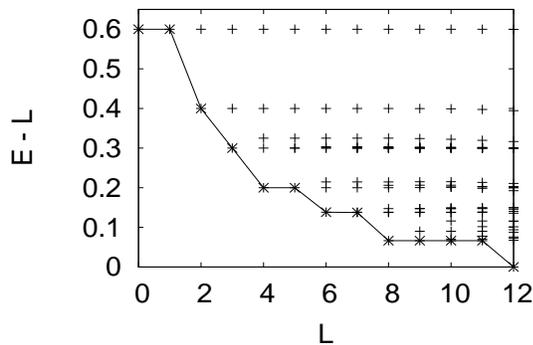}
\vskip2pc
\caption{The full energy spectrum of the many-body Hamiltonian in the rotating frame (in units of $\hbar \omega$), 
for $\Omega = 1$, in a harmonic potential, for $g = 0.1$, $N = 4$ atoms and $0 \le L \le 12$, where $L$ is measured
in units of $\hbar$.}
\end{figure}

\section{Model}

We consider bosonic atoms which are confined in a plane, via a very tight potential in the perpendicular direction, 
and are also subject to an axially-symmetric trapping potential along their plane of motion, $V(\rho)$, where $\rho$ 
is the radial coordinate,
\begin{eqnarray}
 V(\rho) = \frac 1 2 \rho^2 (1 + \lambda \rho^2).
 \label{trpot}
\end{eqnarray} 
This trapping potential is assumed to be either harmonic, $\lambda = 0$, or weakly anharmonic, $0 < \lambda 
\ll 1$. (We set the atom mass $M$, the trap frequency of the harmonic potential $\omega$, and $\hbar$ equal 
to unity). The atom-atom interaction is modelled as the usual contact potential, $V_{\rm int} = g_{2D} 
\delta(\vec{\rho} - \vec{\rho'})$, where $g_{2D}$ is the strength of the effective two-body interaction (for 
the effectively two-dimensional problem that we consider). 

In our analysis below we consider weak interatomic interactions, i.e., we assume that the interaction energy 
is much smaller than the oscillator quantum of energy. This assumption allows us to restrict ourselves to the 
lowest-Landau-level eigenstates of the harmonic potential. 
\begin{equation}
 \psi_m = \frac 1 {\sqrt{\pi m!}} z^m e^{-|z|^2/2},
 \label{LLL}
\end{equation} 
where $z = \rho \exp(i \phi)$, with $\phi$ being the azimuthal angle in cylindrical polar coordinates and
$m \ge 0$ is the quantum number corresponding to the angular momentum (negative values of $m$ correspond 
to states outside the lowest-Landau-level). We stress that even in the case of the anharmonic potential, 
$\lambda \neq 0$, to lowest order in $\lambda$ the eigenstates are still the ones of the harmonic potential.
The corresponding eigenenergies are
\begin{equation}
 \epsilon_m = m + \frac {\lambda} 2 (m+1) (m+2),
 \label{LLLen}
\end{equation} 
where $m \ge 0$.

\begin{figure}[t]
\includegraphics[width=7.5cm,height=5cm,angle=0]{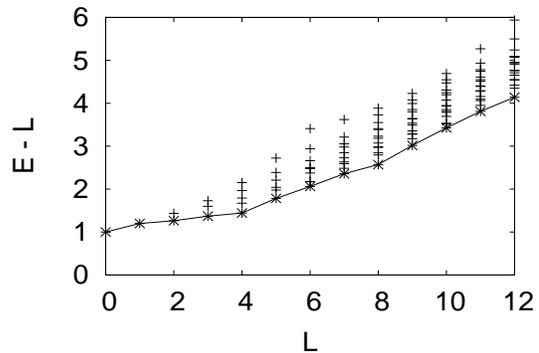}
\vskip2pc
\caption{The full energy spectrum of the many-body Hamiltonian in the rotating frame (in units of $\hbar \omega$), 
for $\Omega = 1$, in an anharmonic potential, for $\lambda = 0.1$, $g = 0.1$, $N = 4$ atoms and $0 \le L \le 12$, 
where $L$ is measured in units of $\hbar$.}
\end{figure}

\subsection{Diagonalization of the many-body Hamiltonian}

The Hamiltonian in second-quantized form is
\begin{eqnarray}
 H = \sum_m \epsilon_m a_m^{\dagger} a_m + \frac g 2 \sum_{m,n,k,l} I_{mnkl} \, 
 a_m^{\dagger} a_n^{\dagger} a_k a_l \, 
 \delta_{m+n, k+l},
\end{eqnarray}
where $a_m$ destroys an atom in the single-particle state $\psi_m$, $g = g_{2D} \int |\psi_0|^4 \, d^2 \rho 
= g_{2D}/(2 \pi)$, and $I_{mnkl} = {(m+n)!}/ [{2^{m+n} \sqrt{m! n! k! l!}}]$. 

Within the lowest-Landau-level approximation there is a degeneracy of all the states with some given angular 
momentum. As a result, one has to use degenerate perturbation theory and diagonalize the interaction term 
within the subspace of these degenerate states. Within this approach all the states with a given atom number 
$N$ and a given angular momentum $L$ are constructed and then the resulting Hamiltonian matrix is diagonalized. 
Obviously one has to truncate the single-particle basis states $\psi_m$ up to some maximum value of the quantum 
number $m$, $m_{\rm max}$.  

\begin{figure}[h]
\includegraphics[width=7.5cm,height=5cm,angle=0]{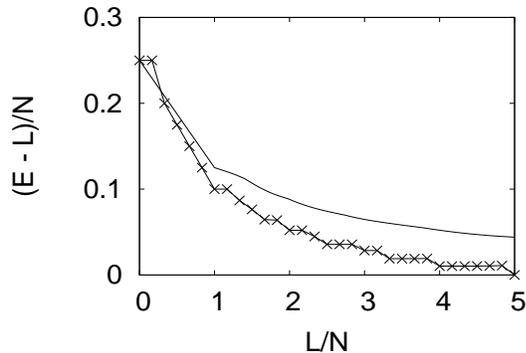}
\vskip2pc
\caption{The (minimum) interaction energy per particle in the rotating frame, $(E - L \Omega)/N$ (in units
of $\hbar \omega)$, with $\Omega = 1$, evaluated within the mean-field approximation (solid line) and within 
the diagonalization of the many-body Hamiltonian (solid line with crosses), in a purely harmonic potential, 
for $g = 0.1$, $N = 6$ atoms and $0 \le L \le 30$, where $L$ is measured in units of $\hbar$.}
\end{figure}
\begin{figure}[h]
\includegraphics[width=7.5cm,height=5cm,angle=0]{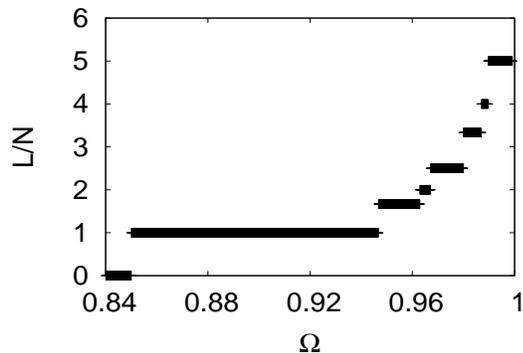}
\vskip2pc
\caption{The function $\ell(\Omega)$ (where $\ell$ is measured in units of $\hbar$ and $\Omega$ in units of $\omega$), 
for $N = 6$ atoms and $g = 0.1$ in a purely harmonic potential, from the energy evaluated within the diagonalization 
of the many-body Hamiltonian of Fig.\,3. The first critical frequency is equal to $0.85$, in agreement with 
Eq.\,(\ref{thres1}).}
\end{figure}
\begin{figure}[h]
\includegraphics[width=7.5cm,height=5cm,angle=0]{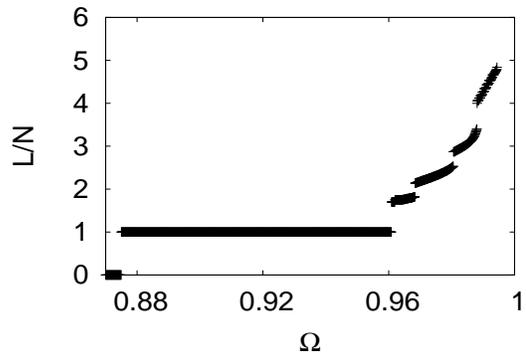}
\vskip2pc
\caption{The function $\ell(\Omega)$ (where $\ell$ is measured in units of $\hbar$ and $\Omega$ in units of $\omega$), 
for $N = 6$ atoms and $g = 0.1$ in a purely harmonic potential, from the energy evaluated within the mean-field 
approximation of Fig.\,3. The first critical frequency is equal to $0.875$, in agreement with Eq.\,(\ref{thres2}).}
\end{figure}

An example of such a calculation is shown in Fig.\,1, where we have considered $N = 4$ atoms, and $g = 0.1$, 
$0 \le m \le 6$, with $L$ ranging from $L = 0$ up to $L = N (N-1) = 12$. [The Laughlin state includes single-particle
states with $m \le 2(N-1)$]. The whole spectrum (i.e., all the eigenvalues) that results from the diagonalization 
of the many-body Hamiltonian is shown in this plot, in the rotating frame, $E - L \Omega$, with $\Omega = 1$.  
In a harmonic potential $E - L$ is also equal with the interaction energy. We observe that the lowest energy 
decreases with increasing $L$, since the gas expands radially. We also see that the lowest energy vanishes for 
$L = 12$, which corresponds to the bosonic Laughlin state. 

The same calculation in an anharmonic potential, with $\lambda = 0.1$ and for the same parameters as in 
Fig.\,1, is shown in Fig.\,2. Contrary to the case of harmonic trapping, here the lowest energies develop
a quasi-periodic behaviour. This is due to the effective potential, which takes the form of a Mexican hat
potential and Bloch's theorem \cite{Bloch}, which implies that (in the case of a ring potential) there is 
a quasi-periodic behaviour of the energy spectrum.

\subsection{Mean-field approximation}

Within the mean-field Gross-Pitaevskii approximation one has to minimize the energy 
\begin{eqnarray}
 E_{\rm mf}/N = - \frac {\hbar^2} {2 M} \int (\Psi_{\ell}^{MF})^* \nabla^2 \Psi_{\ell}^{MF} \, d^2 \rho +
\nonumber \\ 
+ \frac {g_{2D}} 2 (N-1) \int |\Psi_{\ell}^{MF}|^4 \, d^2\rho, 
\end{eqnarray}
where $\Psi_{\ell}^{MF}$ is the mean-field order parameter, with $\ell = L/N$ being the expectation value of 
the angular momentum per atom. In the case of harmonic confinement only the interaction energy, i.e., the second 
term has to be minimized, since for some fixed value of the angular momentum the first term is equal to $L$. 

A convenient way to do that is to expand the order parameter in the basis of the single-particle eigenstates
$\psi_m$,
\begin{eqnarray}
 \Psi_{\ell}^{MF} = \sum_{m=0}^{m_{\rm max}} c_m \psi_m
 \label{expan}
\end{eqnarray}
and then perform the minimization under the two constraints of particle normalization $\sum_{m} c_m^2 = 1$
and fixed angular momentum $\sum_{m} m \, c_m^2 = L/N = \ell$.

\section{Energy and rotational response} 

\subsection{Harmonic trapping potential}

Following the two approaches described above we plot in Fig.\,3 the result from the minimization of the energy 
for $N=6$ and $g = 0.1$ evaluated within the mean-field approximation, for $0 \le \ell \le 5$. What is actually
plotted is the energy in the rotating frame $(E - L \Omega)/N$, with $\Omega = 1$. (As mentioned also above, this  
coincides with the interaction energy in a purely harmonic potential.) 

In the same figure, we also plot the lowest-energy eigenvalue of the many-body Hamiltonian for $N = 6$ and $0 \le 
L \le 30$. We see that the energy that is evaluated from the diagonalization is lower than that of the mean-field, 
as expected. The only exception is the case $L=1$. This state has the same energy as the ground state with $L = 0$ 
due to the center of mass excitation. 

While for ``small" values of $\ell$ the deviation between the two approaches is correspondingly ``small" (i.e., of
order $N$), as $L$ approaches $30$ [i.e., $L = N(N-1)]$, the energy difference increases, becoming of order $N^2$. 
We also observe that in the range $N(N-2) \le L \le N(N-1)-1$ for these $N-1$ values of $L$ the energies are 
degenerate and result from exciting the center of mass motion of the state with $L = N(N-2)$. Similar plateaus 
also form for smaller values of $L$.

Up to now we have minimized the energy of the system $E(\ell)$ for some fixed angular momentum $\ell$. 
Another interesting, and physically-relevant calculation is the one where the rotational frequency of the trapping 
potential $\Omega$ is fixed, instead. Making the implicit assumption that the trap has an infinitesimal asymmetry 
and that the cloud equilibrates in the rotating frame, one has to minimize the energy in the rotational frame, 
$E(\ell) - \ell \Omega$, for some fixed value of $\Omega$. Following this procedure the function $\ell(\Omega)$ 
may be derived.

\begin{figure}[t]
\includegraphics[width=7.5cm,height=5cm,angle=0]{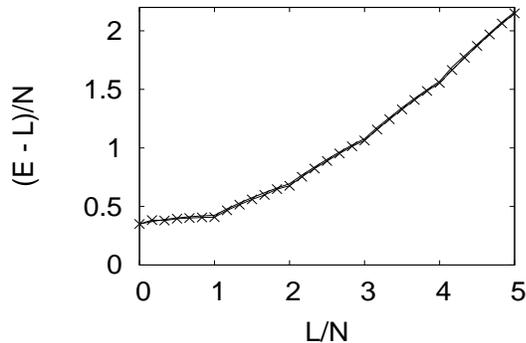}
\vskip2pc
\caption{The (minimum) energy per particle in the rotating frame, $(E - L \Omega)/N$ (in units of $\hbar \omega)$, 
with $\Omega = 1$, evaluated within the mean-field approximation (solid line) and within the diagonalization of the 
many-body Hamiltonian (solid line with crosses), in an anharmonic potential, with $\lambda = 0.1$ and $g = 0.1$, 
$N = 6$ atoms and $0 \le L \le 30$, where $L$ is measured in units of $\hbar$. The difference between the two curves 
is hardly visible.} 
\end{figure}
\begin{figure}[h]
\includegraphics[width=7.5cm,height=5cm,angle=0]{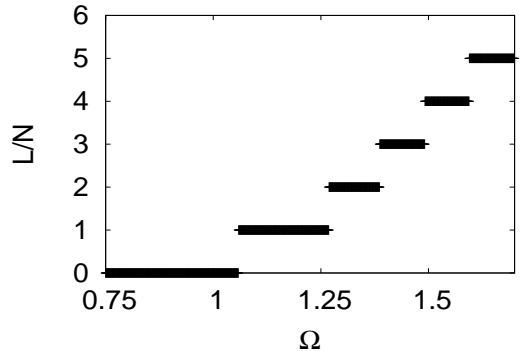}
\vskip2pc
\caption{The function $\ell(\Omega)$ (where $\ell$ is measured in units of $\hbar$ and $\Omega$ in units of $\omega$), 
for $N = 6$ atoms and $g = 0.1$ in a anharmonic potential, with $\lambda = 0.1$ from the energy evaluated within the 
diagonalization of the many-body Hamiltonian of Fig.\,6. The first critical frequency is $\approx 1.059$.}
\end{figure}
\begin{figure}[h]
\includegraphics[width=7.5cm,height=5cm,angle=0]{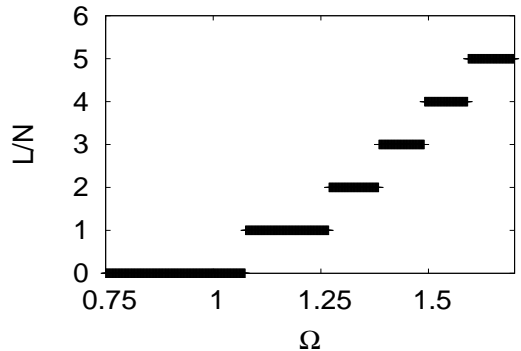}
\vskip2pc
\caption{The function $\ell(\Omega)$ (where $\ell$ is measured in units of $\hbar$ and $\Omega$ in units of $\omega$), 
for $N = 6$ atoms and $g = 0.1$ in a anharmonic potential, with $\lambda = 0.1$ from the energy evaluated within the 
mean-field approximation of Fig.\,6. The first critical frequency is equal to $1.075$, in agreement with 
Eq.\,(\ref{thres3}).}
\end{figure}

Figure 4 shows $\ell(\Omega)$ for the energy that is evaluated from the diagonalization approach, while Fig.\,5 
from the mean-field approach \cite{Rok}. The first critical frequency $\Omega_1$ associated with a single vortex 
state entering the cloud, evaluated within the diagonalization is given by 
\begin{eqnarray}
 \Omega_1 = 1 - \frac 1 4 N g.
 \label{thres1}
\end{eqnarray}
The same frequency, evaluated within the mean-field approximation, is
\begin{eqnarray}
 \Omega_1 = 1 - \frac 1 4 (N-1) g.
 \label{thres2}
\end{eqnarray}
For this reason $\Omega_1$ is 0.85 in Fig.\,4, and 0.875 in Fig.\,5. As we observe, the finiteness of $N$
decreases the value of $\Omega_1$.

\subsection{Anharmonic trapping potential}

Turning to the case of anharmonic confinement, we present the same results as in the previous subsection, for the
same parameters and with $\lambda = 0.1$. Figure 6 shows the energy evaluated within the two approaches, i.e., the
diagonalization and the mean-field. The difference between the two is hardly visible. 

The function $\ell(\Omega)$ that results from the energies of Fig.\,6, is shown in Figs.\,7 and 8. One major 
difference between the present, anharmonic, confinement and the harmonic one is that $\Omega$ may exceed unity in 
this case. The critical frequency for the formation of one vortex state is, evaluated within the mean-field 
approximation  
\begin{eqnarray}
 \Omega_1 = 1 + 2 \lambda - \frac 1 4 (N-1) g.
 \label{thres3}
\end{eqnarray}
Indeed, the value of $\Omega_1$ is 1.075 in Fig.\,8, while $\Omega_1$ is slightly smaller, $\approx 1.059$, in
Fig.\,7. As in the case of harmonic confinement $\Omega_1$ is smaller in a system with a small atom number, however
here the difference between the two frequencies is much smaller.

In addition, in the present case of anharmonic confinement $\ell(\Omega)$ consists of rather regular steps, as a 
result of the quasi-periodic behaviour of the dispersion relation.

\section{Single-particle density distribution and the pair-correlation function}

We turn now to the single-particle density distribution $n(\vec{\rho})$, and the the pair-correlation 
function $g^{(2)}(\vec{\rho}, \vec{\rho'})$. The single-particle density is simply
\begin{eqnarray}
  n(\vec{\rho}) = \langle \Phi^{\dagger}(\vec{\rho}) \Phi(\vec{\rho}) \rangle,
\end{eqnarray} 
where $\Phi(\vec{\rho})$ is the operator that destroys a particle at $\vec{\rho}$. Because of the axial symmetry of
the problem $n(\vec{\rho})$ is also axially symmetric, $n(\vec{\rho})=n(\rho)$. 

Also, the pair-correlation function is
\begin{eqnarray}
 g^{(2)}(\vec{\rho}, \vec{\rho'}) = \frac {\langle \Phi^{\dagger}(\vec{\rho}) \Phi^{\dagger}(\vec{\rho'}) 
 \Phi(\vec{\rho'}) \Phi(\vec{\rho}) \rangle} 
 {\langle \Phi^{\dagger}(\vec{\rho}) \Phi(\vec{\rho}) \rangle 
 \langle \Phi^{\dagger}(\vec{\rho'}) \Phi(\vec{\rho'}) \rangle}. 
\end{eqnarray}
In the calculations which are shown in the plots, $\vec{\rho}$ and $\vec{\rho'}$ are assumed to point in the same
direction.

For $L=0$, where the many-body state is simply the uncorrelated, mean-field state $|0^N \rangle$,
$g^{(2)}(\vec{\rho}, \vec{\rho'})$ is a straight line and equal to $(N-1)/N$. Another easy case is
the one with $L=1$, where the lowest-energy state is $|0^{N-1}, L^1 \rangle$. In this case
\begin{eqnarray}
 g^{(2)}(\rho, \rho') = \frac {(N-1) [N-2+(\rho-\rho')^2])} {(N-1+\rho^2) (N-1+{\rho'}^2)}. 
\end{eqnarray} 
For large values of $N$ and finite $\rho$ and $\rho'$, $g^{(2)}(\vec{\rho}, \vec{\rho'}) \to (N-2)/(N-1)$.
This behaviour is very different than the limit where we have the correlated (Laughlin) states. There, 
$g^{(2)}(\vec{\rho}, \vec{\rho'})$ has to vanish for $\vec{\rho} = \vec{\rho'}$ due to the correlations 
between the atoms. As a result $g^{(2)}(\vec{\rho}, \vec{\rho'})$ remains spatially dependent, even for 
large values of $N$. 
 
\begin{figure}[t]
\includegraphics[width=7cm,height=4.6cm,angle=0]{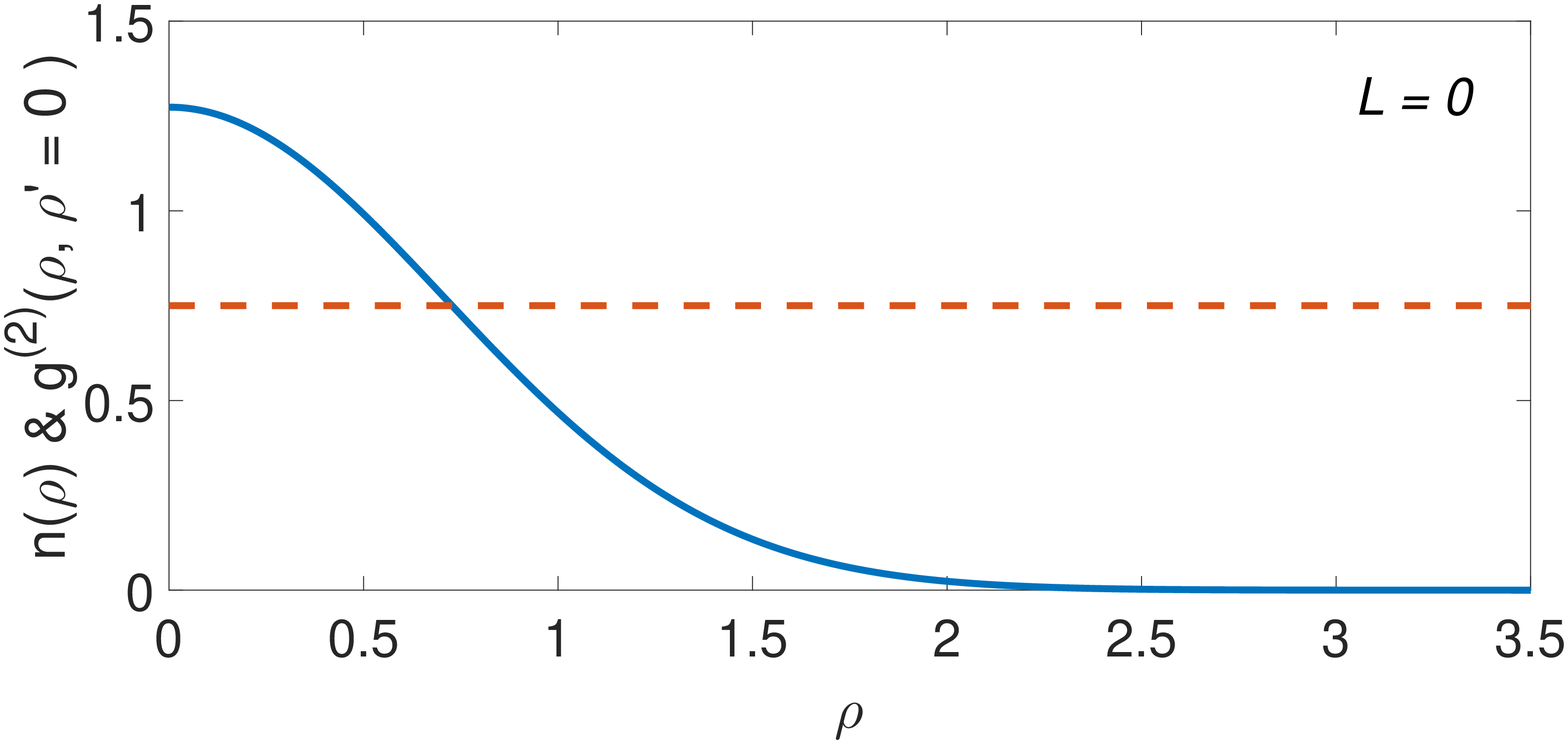}
\includegraphics[width=7cm,height=4.6cm,angle=0]{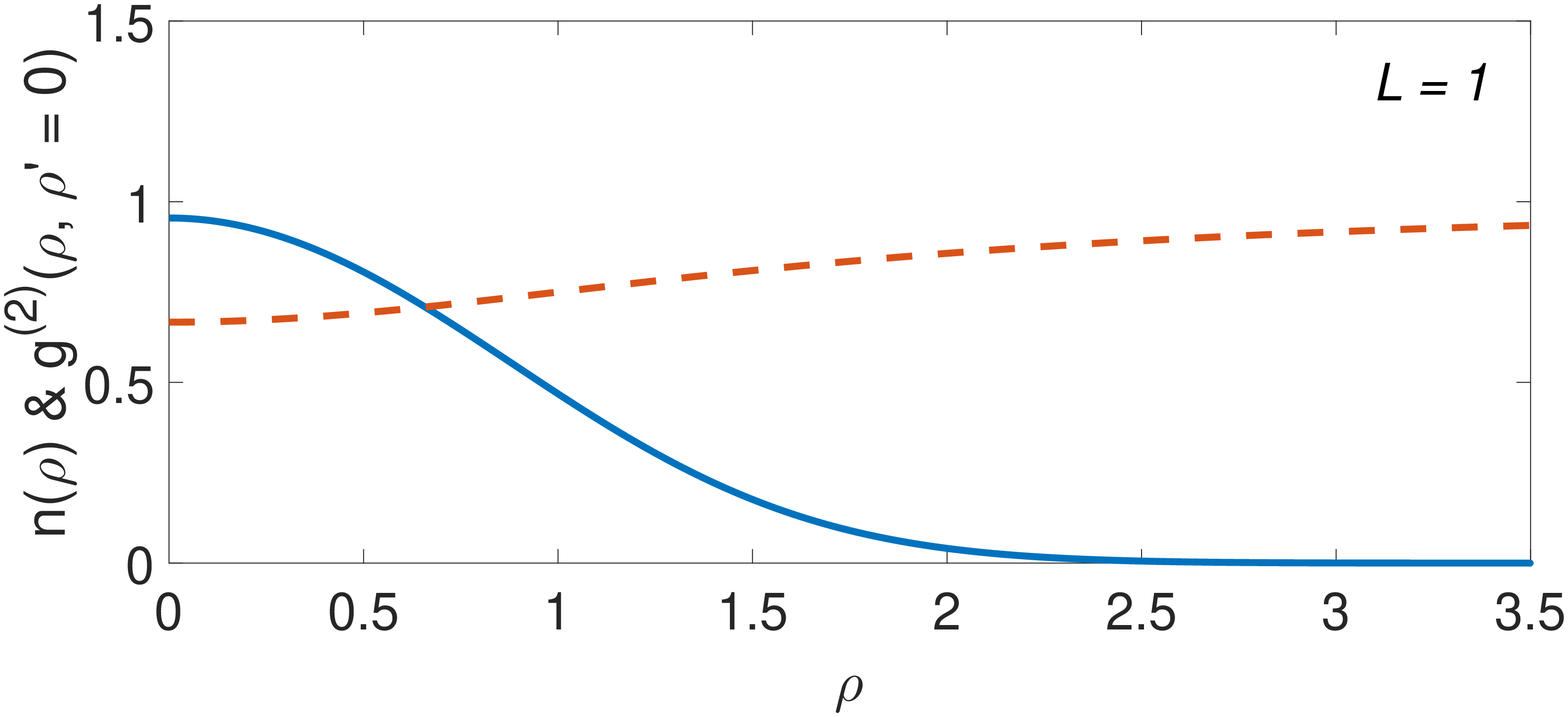}
\includegraphics[width=7cm,height=4.6cm,angle=0]{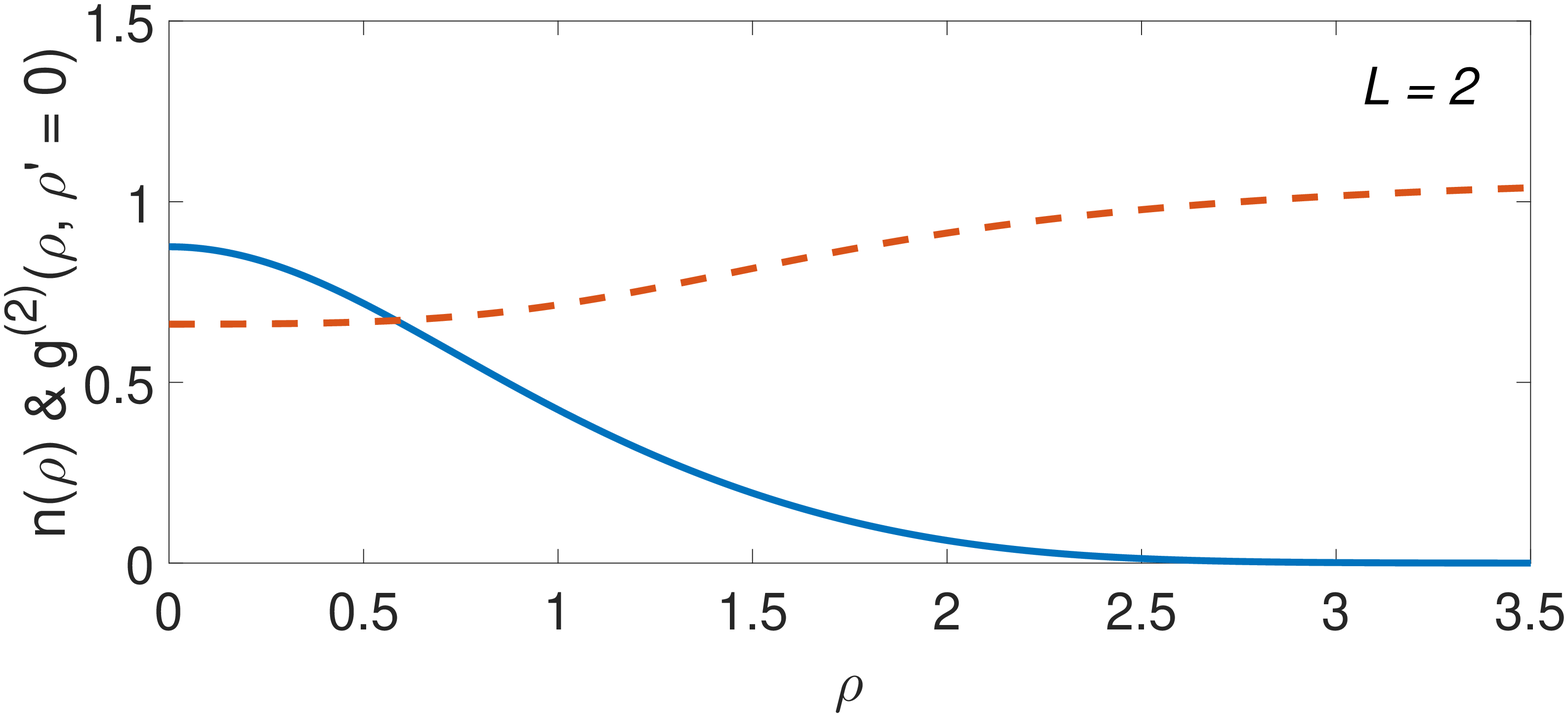}
\includegraphics[width=7cm,height=4.6cm,angle=0]{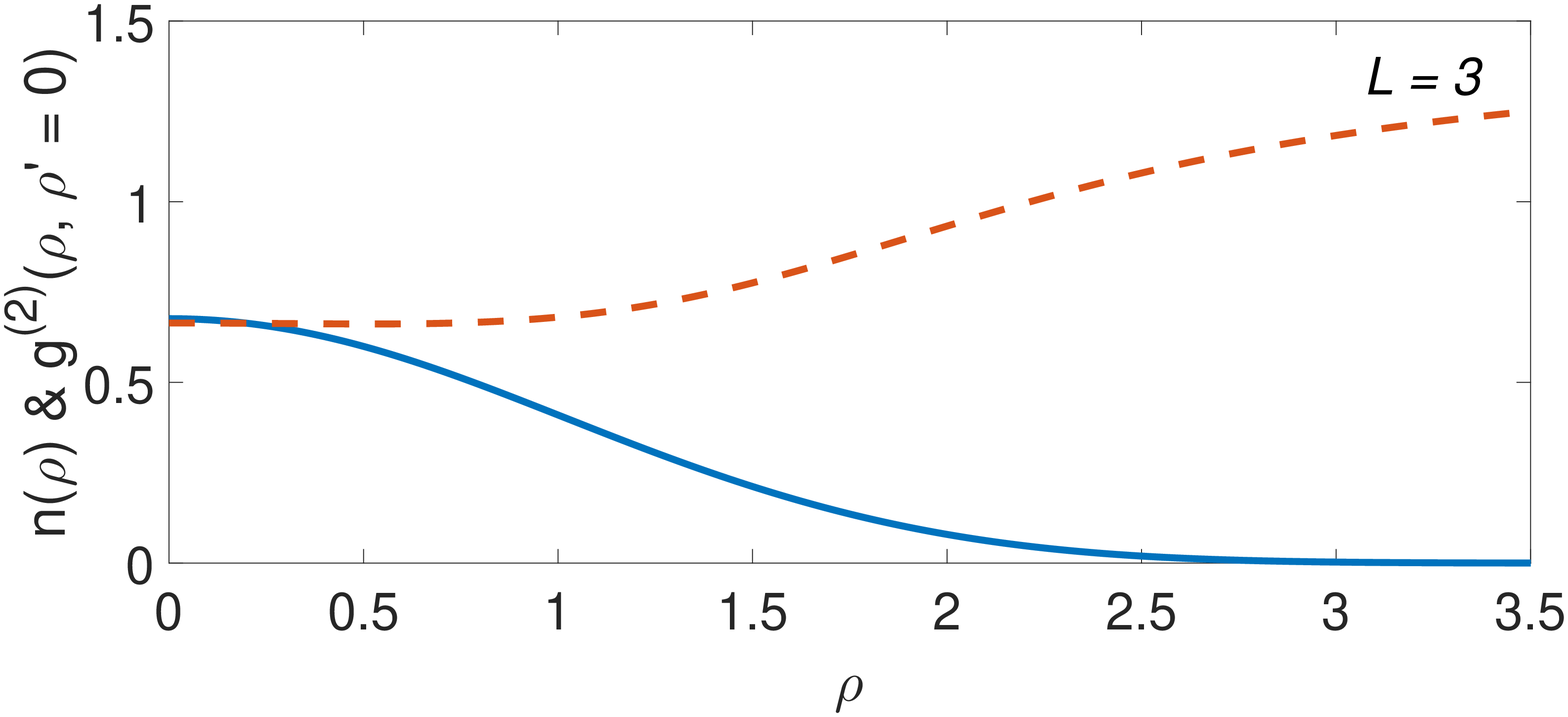}
\caption{(Color online) The density $n(\rho)$ (solid line), in units of $a_0^{-2}$, where $a_0 = \sqrt{\hbar/(M \omega)}$ 
is the oscillator length, and the pair-correlation function $g^{(2)}(\rho, \rho' = 0)$ (dashed line) for the lowest-energy 
eigenstate of the Hamiltonian, for $N = 4$ atoms, and $0 \le L \le 3$, for $g = 0.1$. Also $\rho$ is measured in units
of $a_0$.}
\end{figure}
\begin{figure}[h]
\includegraphics[width=7cm,height=4.6cm,angle=0]{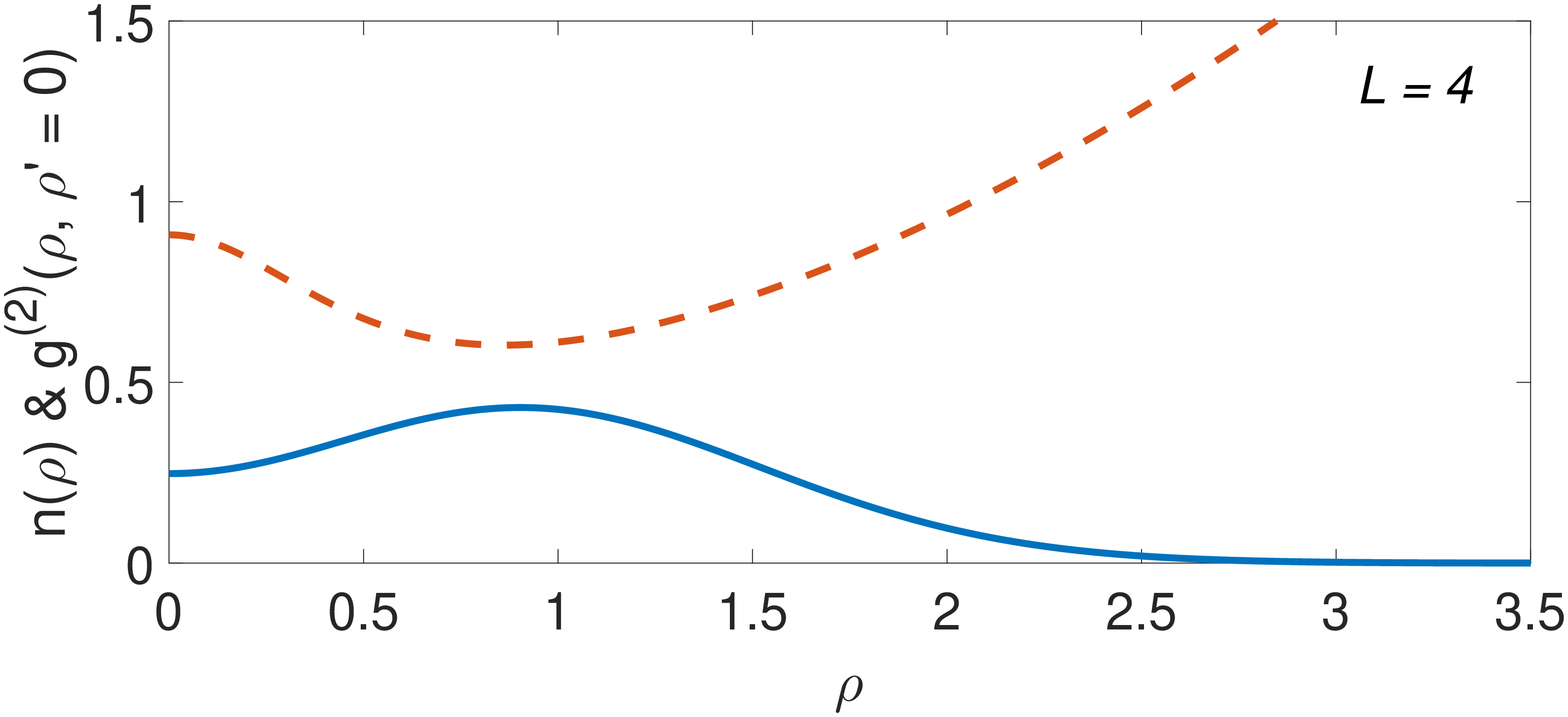}
\includegraphics[width=7cm,height=4.6cm,angle=0]{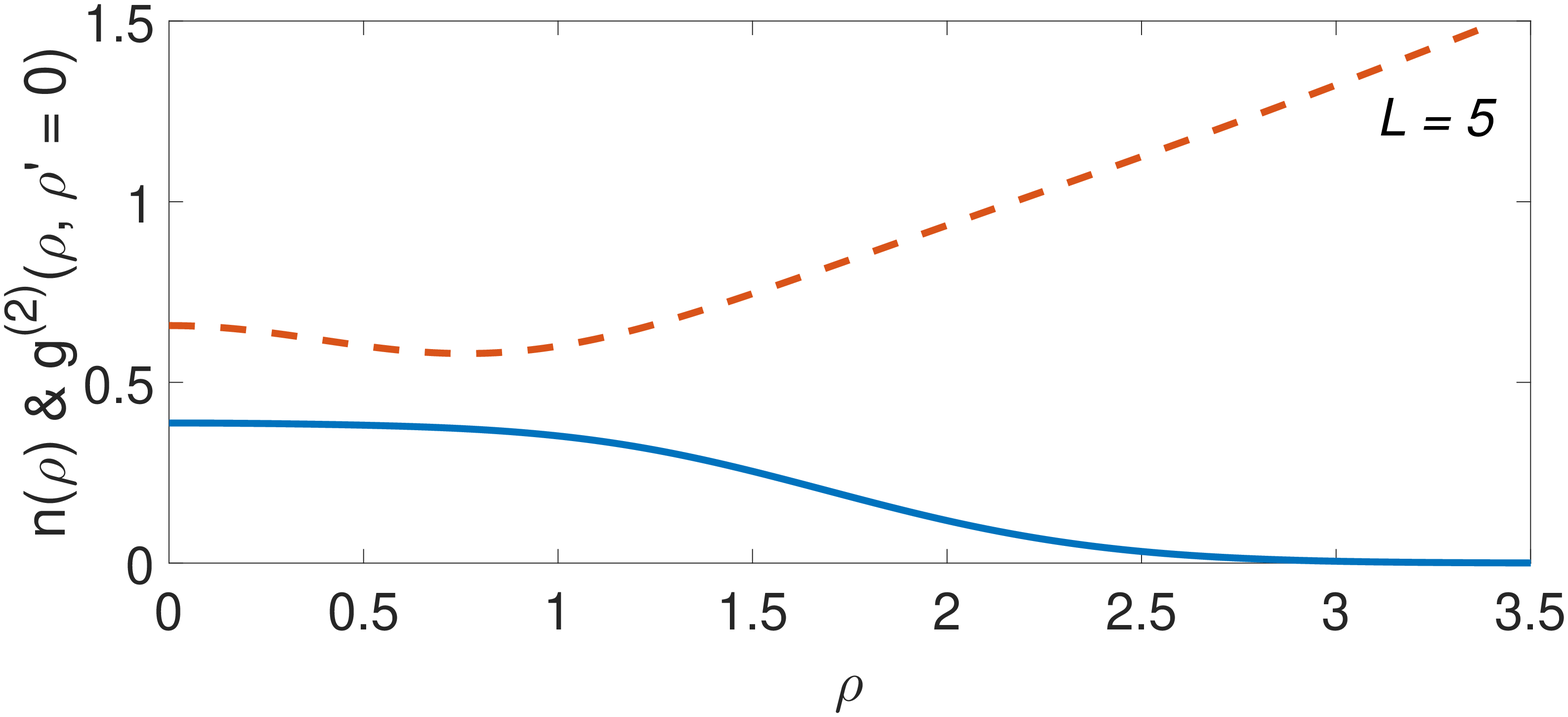}
\includegraphics[width=7cm,height=4.6cm,angle=0]{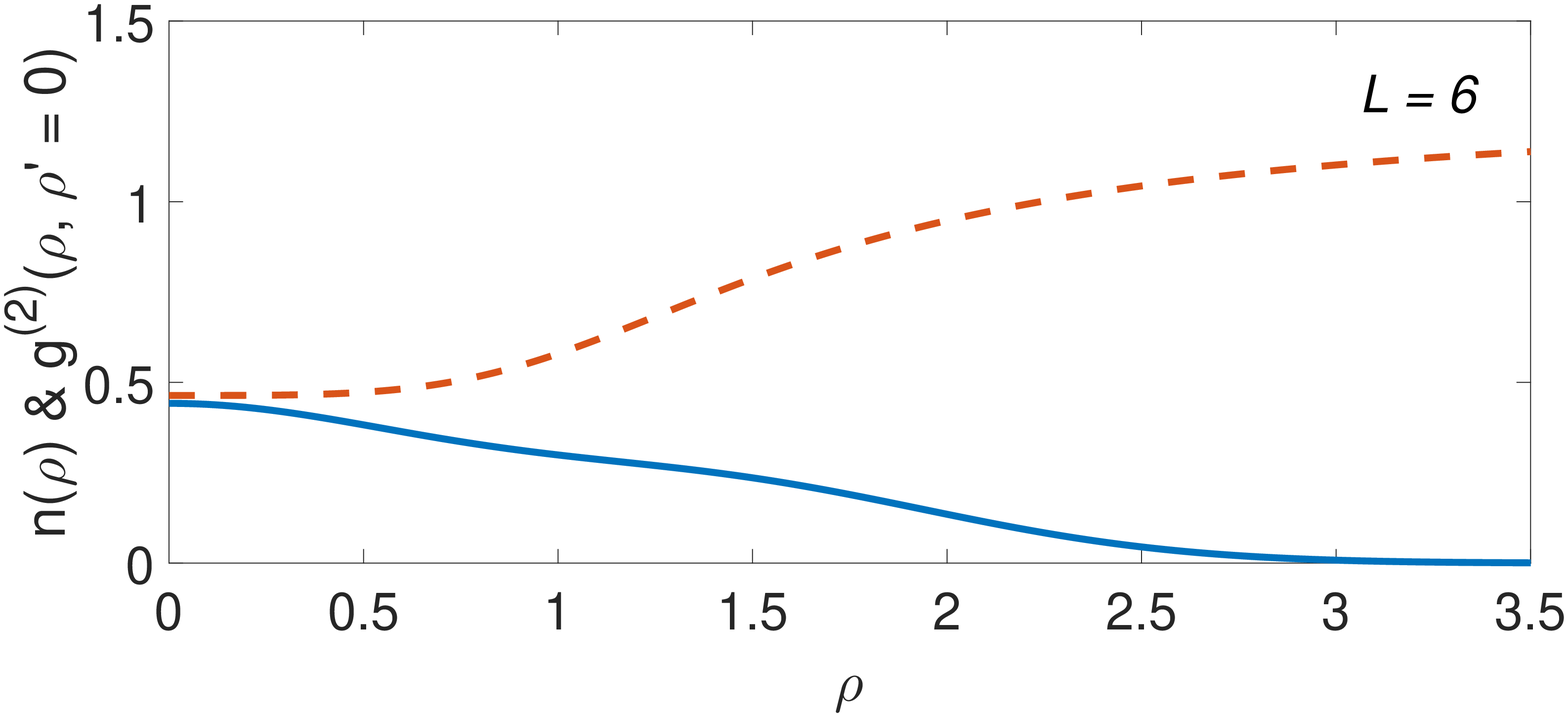}
\includegraphics[width=7cm,height=4.6cm,angle=0]{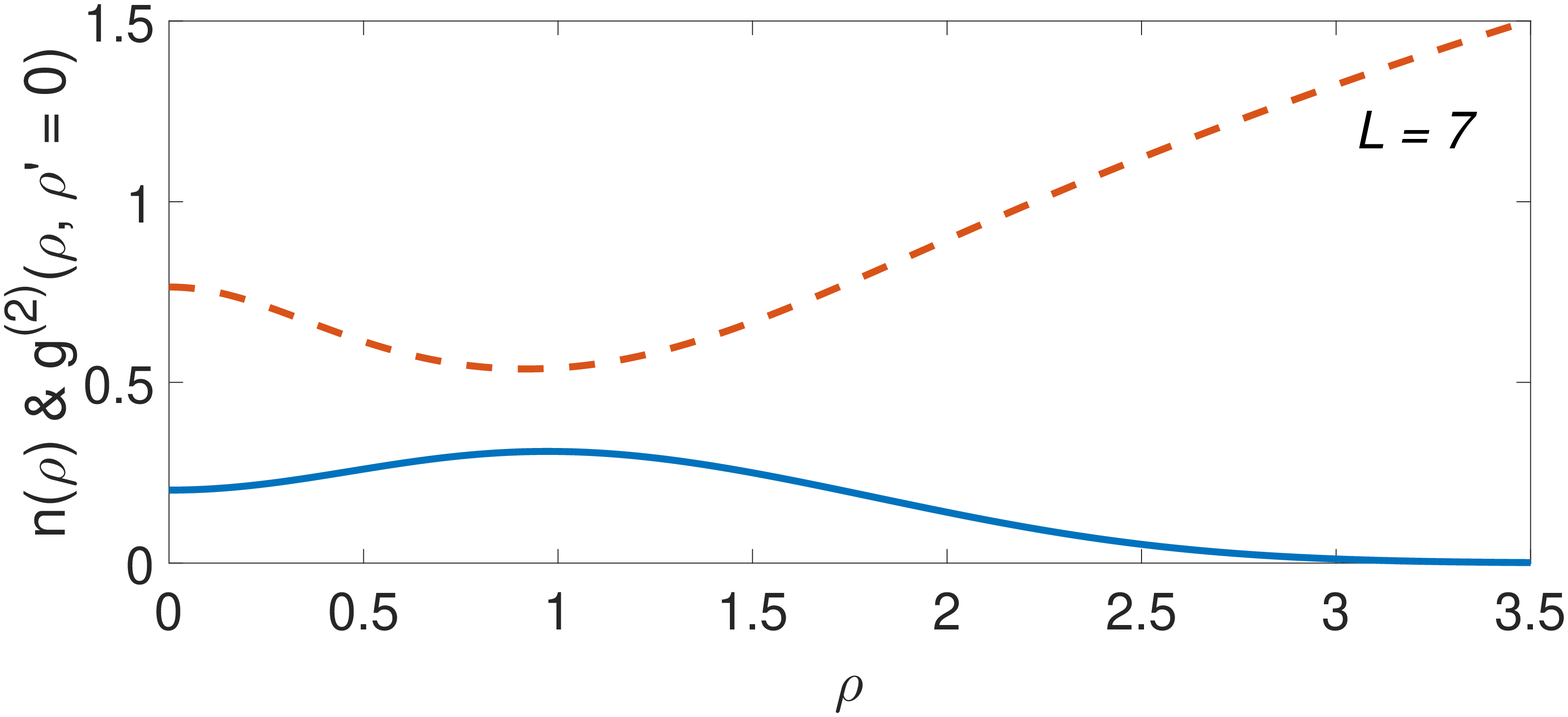}
\caption{(Color online) The density $n(\rho)$ (solid line) in units of $a_0^{-2}$, and the pair-correlation function 
$g^{(2)}(\rho, \rho' = 0)$ (dashed line) for the lowest-energy eigenstate of the Hamiltonian, for $N = 4$ atoms, and 
$4 \le L \le 7$, for $g = 0.1$. Also $\rho$ is measured in units of $a_0$.}
\end{figure}
\begin{figure}[h]
\includegraphics[width=7cm,height=4.6cm,angle=0]{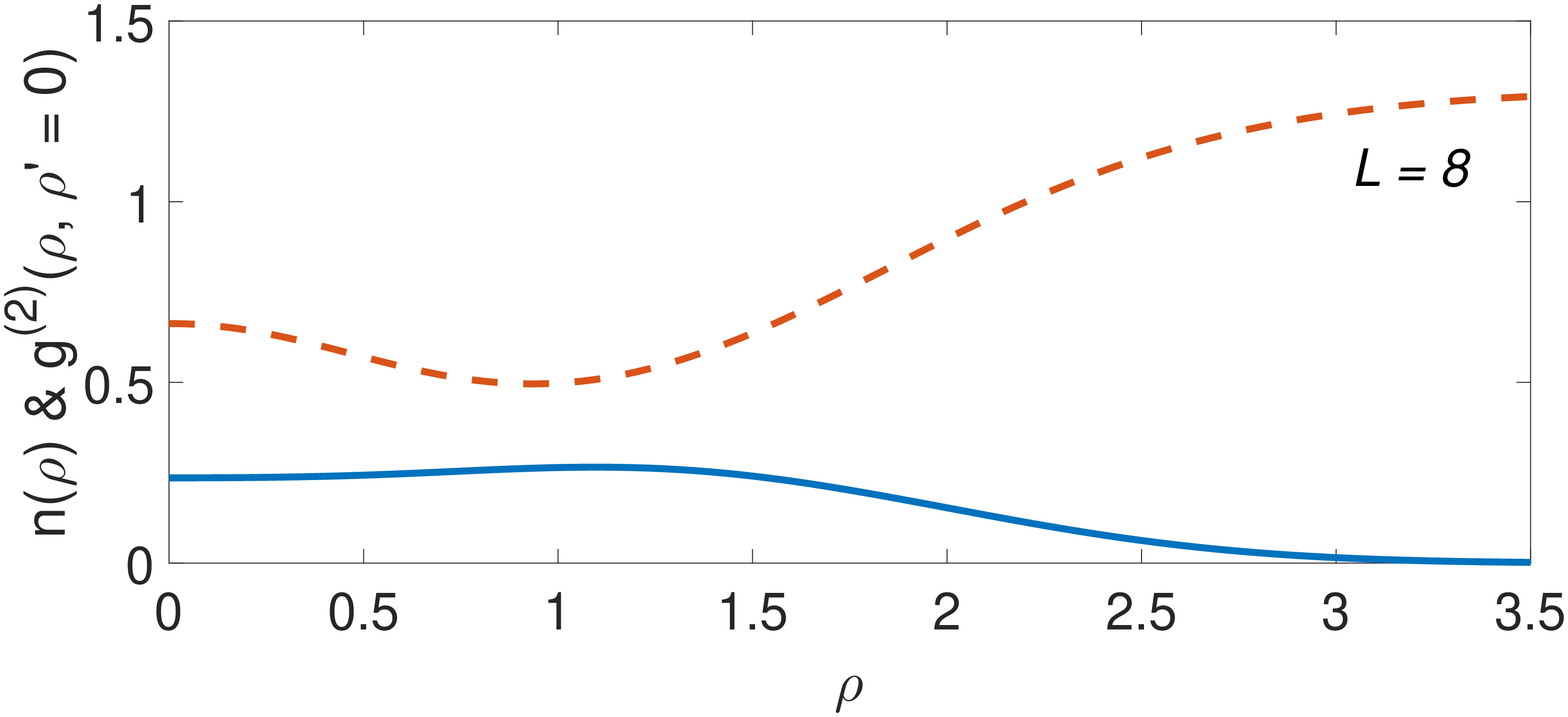}
\includegraphics[width=7cm,height=4.6cm,angle=0]{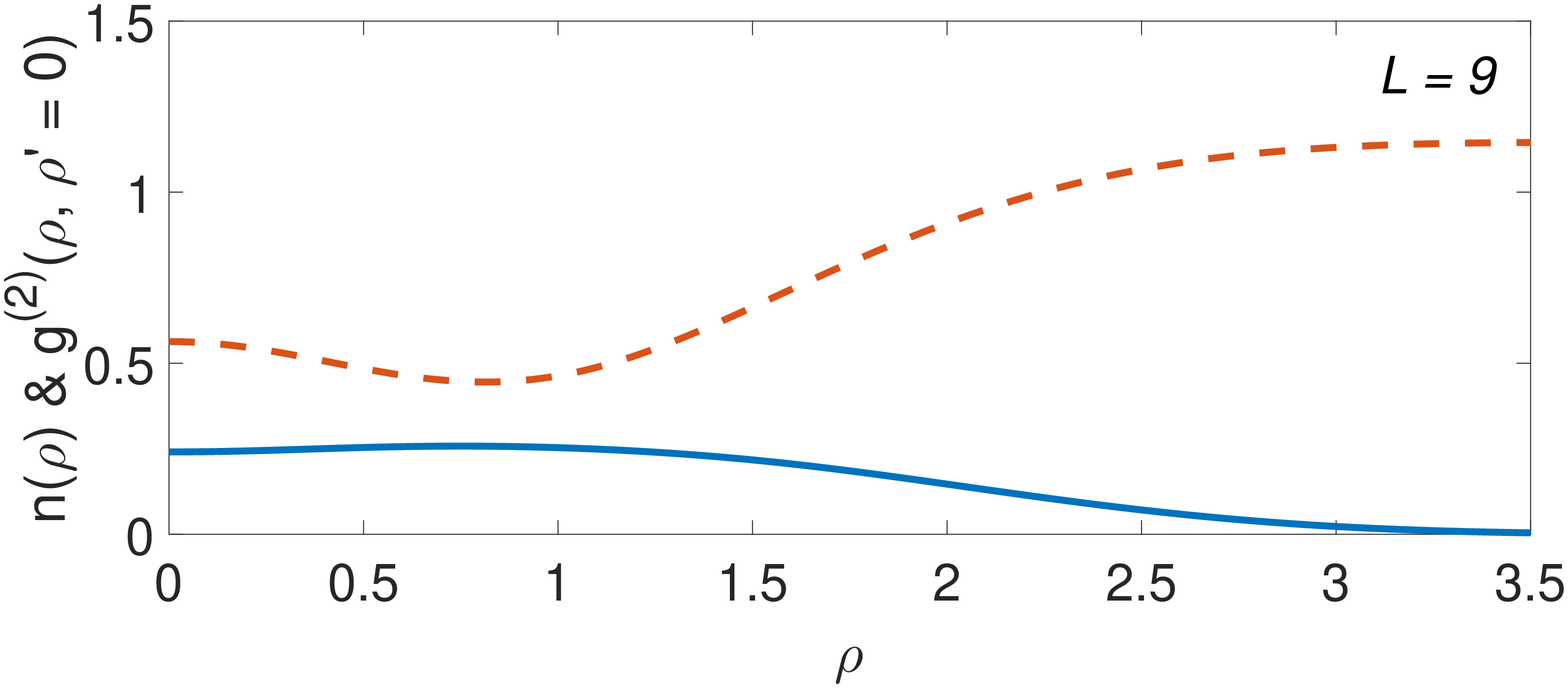}
\includegraphics[width=7cm,height=4.6cm,angle=0]{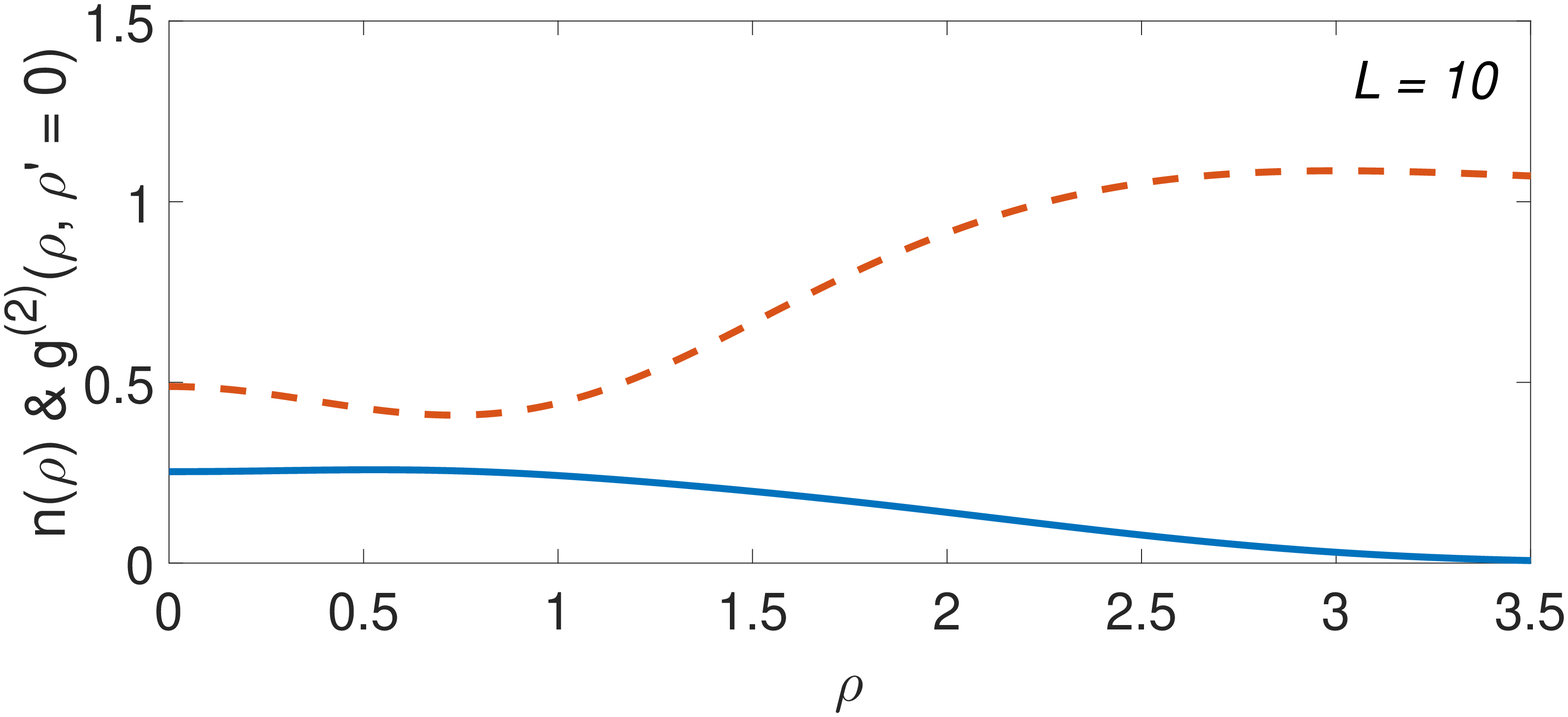}
\includegraphics[width=7cm,height=4.6cm,angle=0]{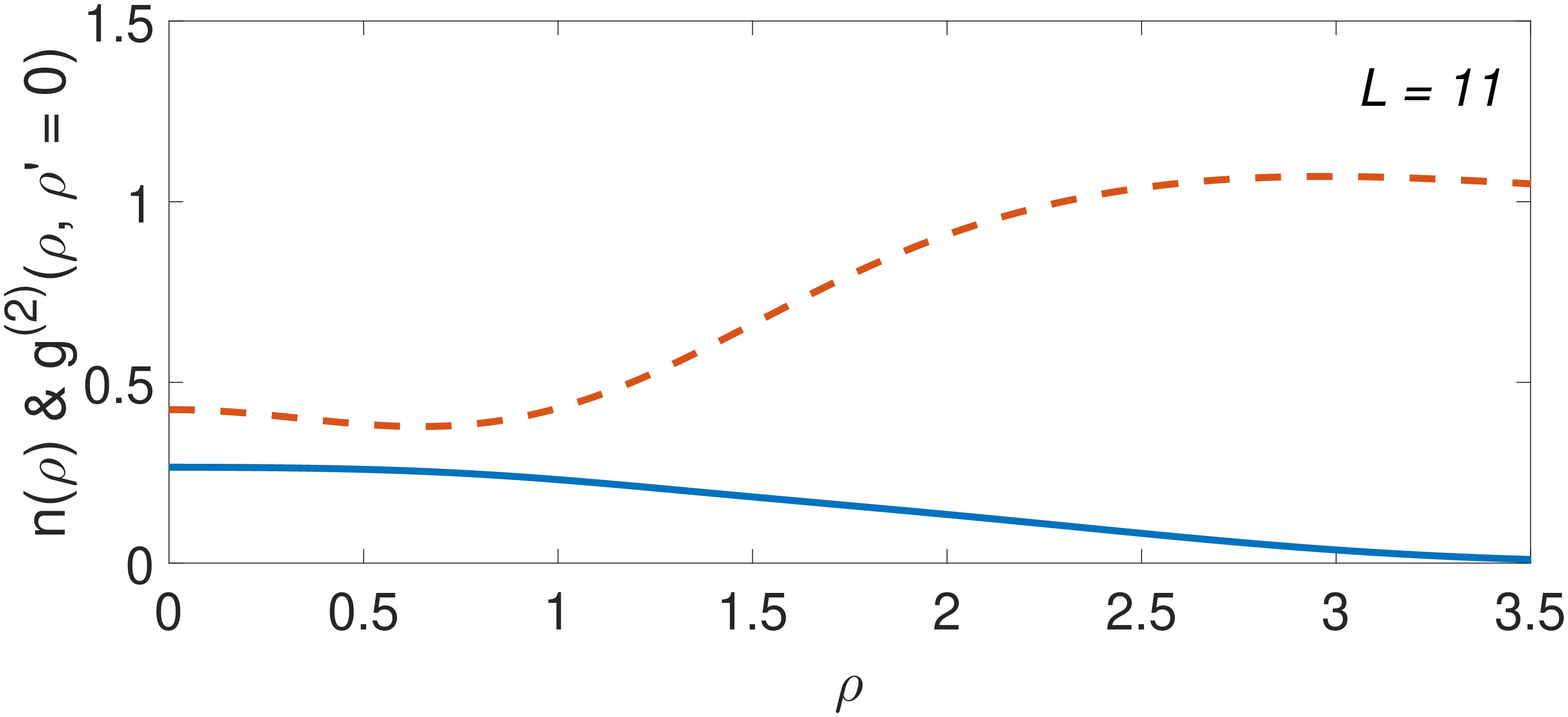}
\caption{(Color online) The density $n(\rho)$ (solid line) in units of $a_0^{-2}$, and the pair-correlation function 
$g^{(2)}(\rho, \rho' = 0)$ (dashed line) for the lowest-energy eigenstate of the Hamiltonian, for $N = 4$ atoms, and 
$8 \le L \le 11$, for $g = 0.1$. Also $\rho$ is measured in units of $a_0$.}
\end{figure}
\begin{figure}[h]
\includegraphics[width=7cm,height=4.6cm,angle=0]{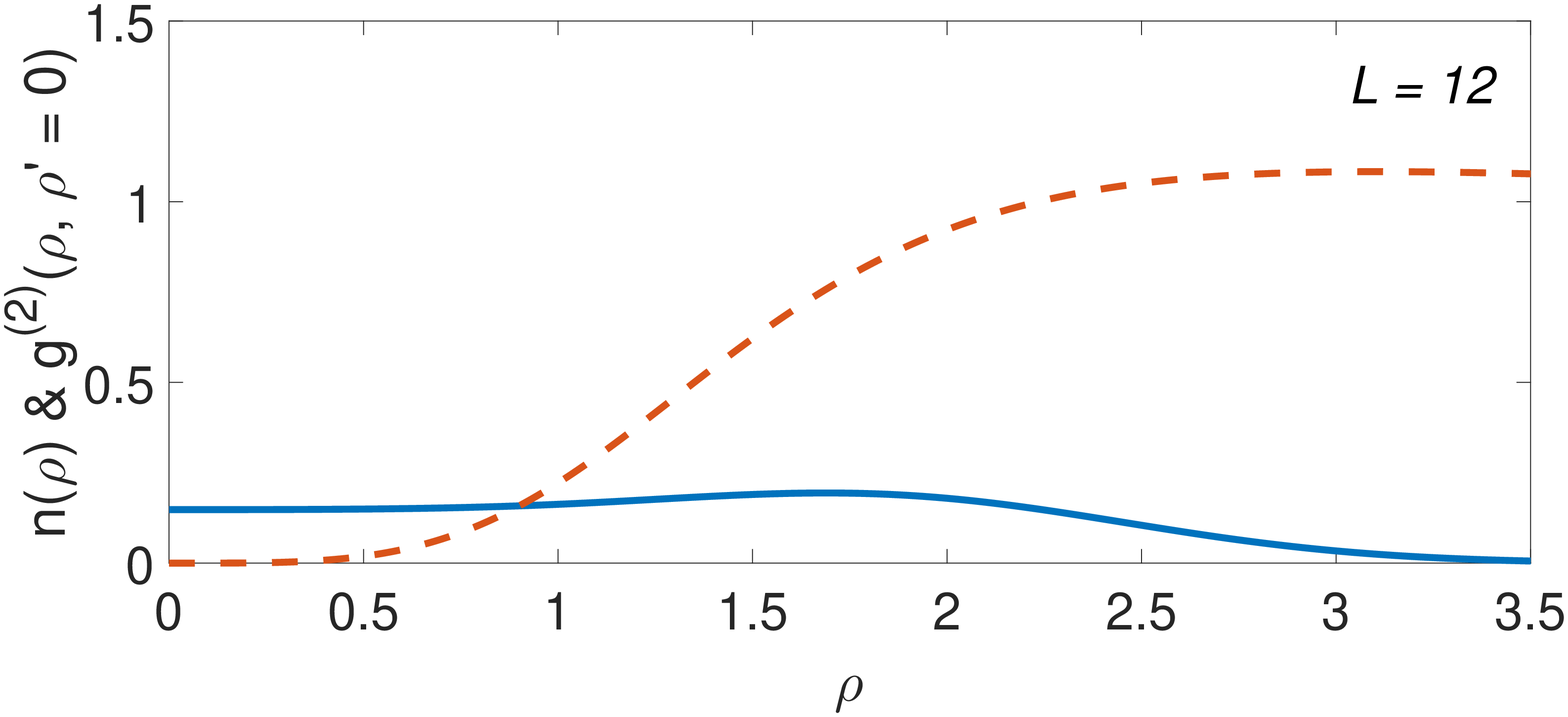}
\caption{(Color online) The density $n(\rho)$ (solid line) in units of $a_0^{-2}$, and the pair-correlation function 
$g^{(2)}(\rho, \rho' = 0)$ (dashed line) for the lowest-energy eigenstate of the Hamiltonian, for $N = 4$ atoms, and 
$L = 12$, i.e., the Laughlin state, for $g = 0.1$. Also $\rho$ is measured in units of $a_0$.}
\end{figure}

\subsection{Harmonic potential}

In Figs.\,9, 10, 11 and 12 we have considered $N = 4$ atoms and we have plotted the single-particle density distribution
$n(\rho)$, as well as the pair-correlation function $g^{(2)}(\rho, \rho')$, with the reference point $\rho'$ located 
at zero. We observe that as the angular momentum increases, $n(\rho)$ becomes more extended because the atoms expand 
radially due to their rotational motion. Furthermore, $n(\rho)$ also becomes more flat (the Laughlin state is the 
last one, with $L=12$).

In these figures we also observe that $g^{(2)}(\rho, \rho' = 0)$ develops a node at $\rho = 0$ only for $L=12$ (i.e., 
for the Laughlin state).

\begin{figure}[t]
\includegraphics[width=7cm,height=4.6cm,angle=0]{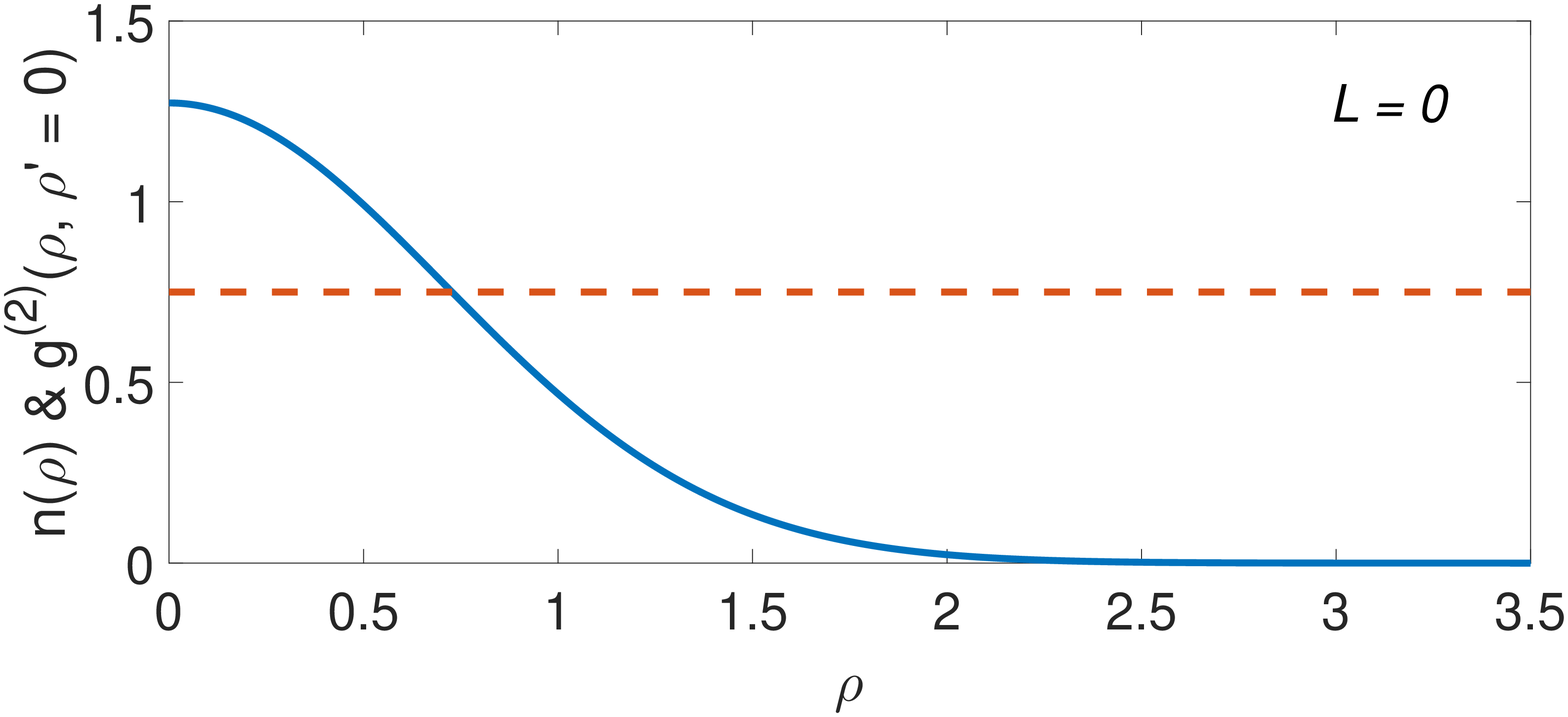}
\includegraphics[width=7cm,height=4.6cm,angle=0]{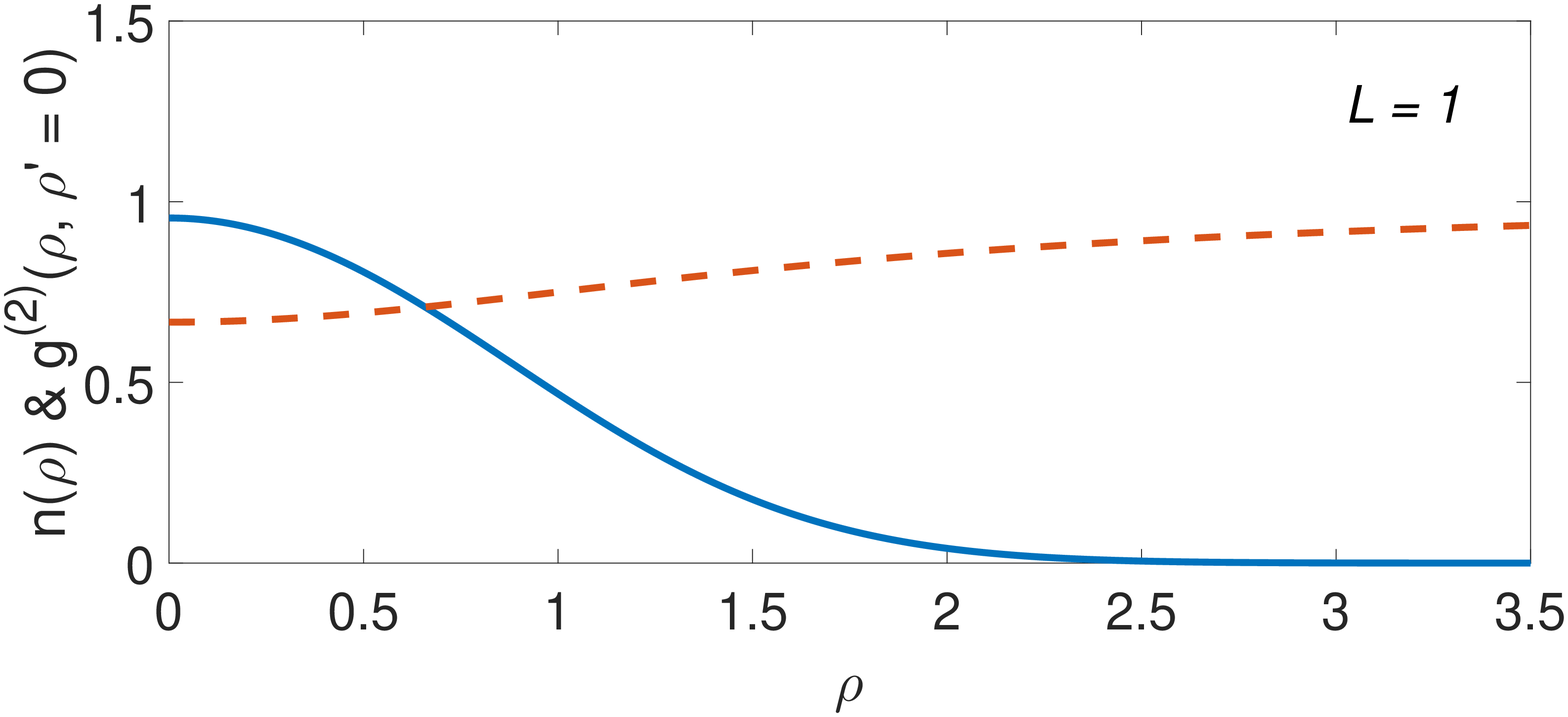}
\includegraphics[width=7cm,height=4.6cm,angle=0]{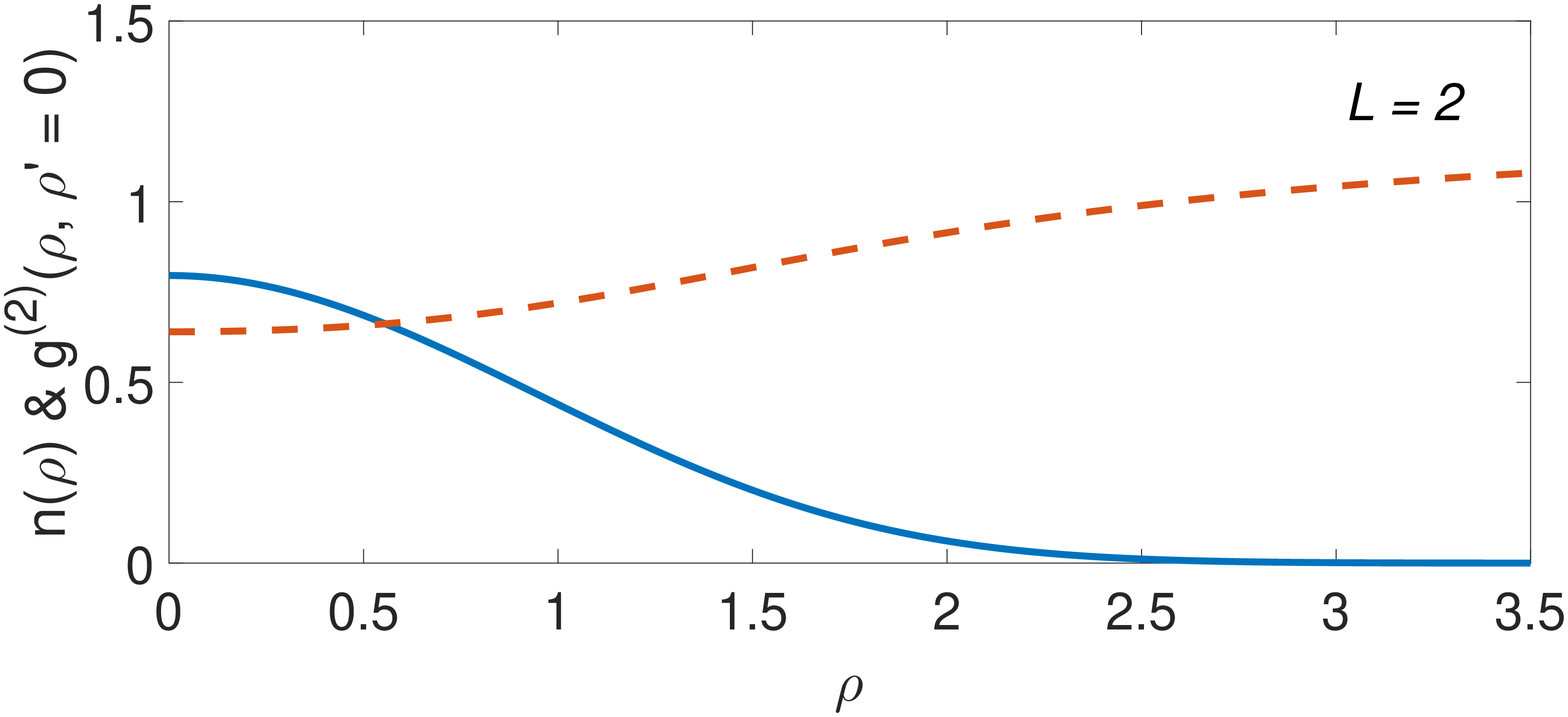}
\includegraphics[width=7cm,height=4.6cm,angle=0]{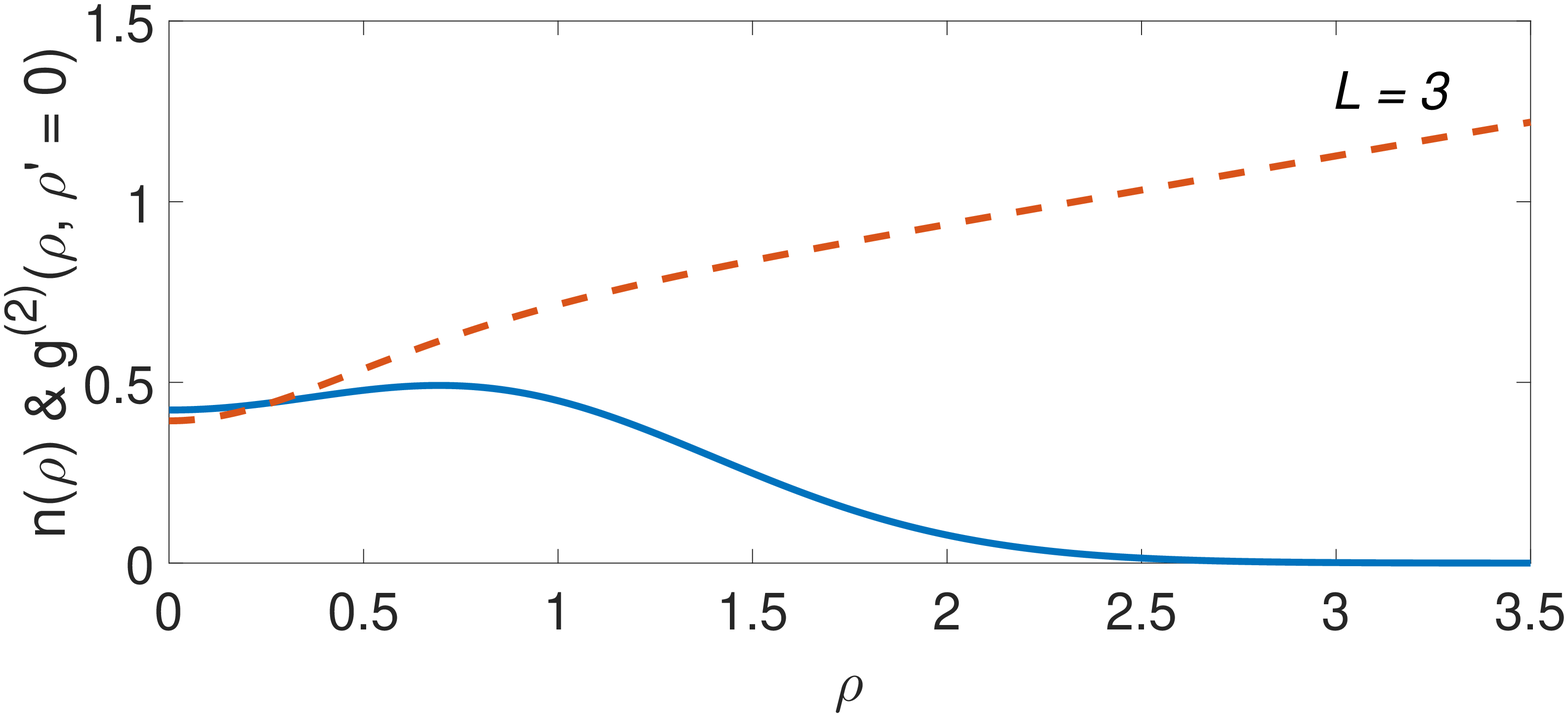}
\caption{(Color online) The density $n(\rho)$ (solid line) in units of $a_0^{-2}$, and the pair-correlation function 
$g^{(2)}(\rho, \rho' = 0)$ (dashed line) for the lowest-energy eigenstate of the Hamiltonian, for $N = 4$ atoms, and 
$0 \le L \le 3$, for $g = 0.1$ and $\lambda = 0.1$. Also $\rho$ is measured in units of $a_0$.}
\end{figure}
\begin{figure}[h]
\includegraphics[width=7cm,height=4.6cm,angle=0]{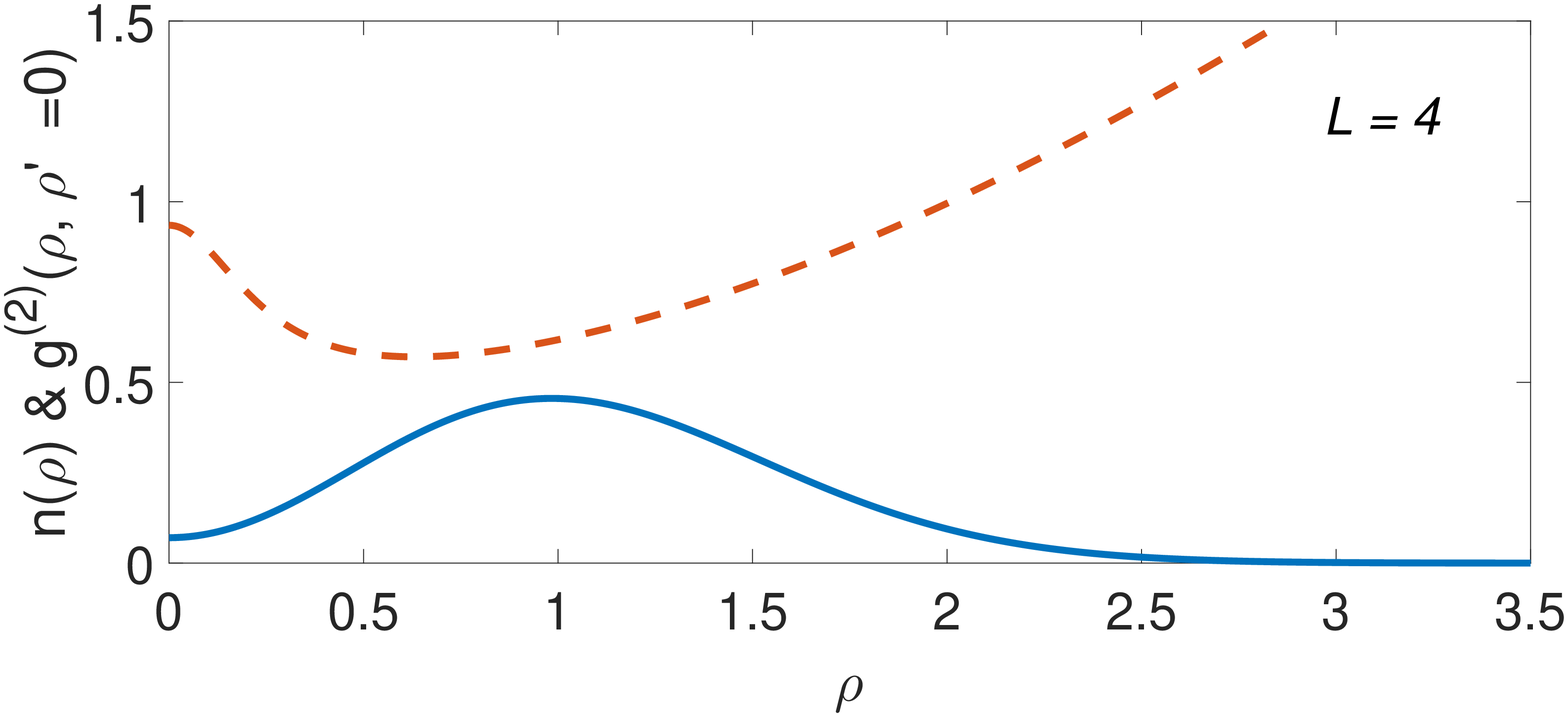}
\includegraphics[width=7cm,height=4.6cm,angle=0]{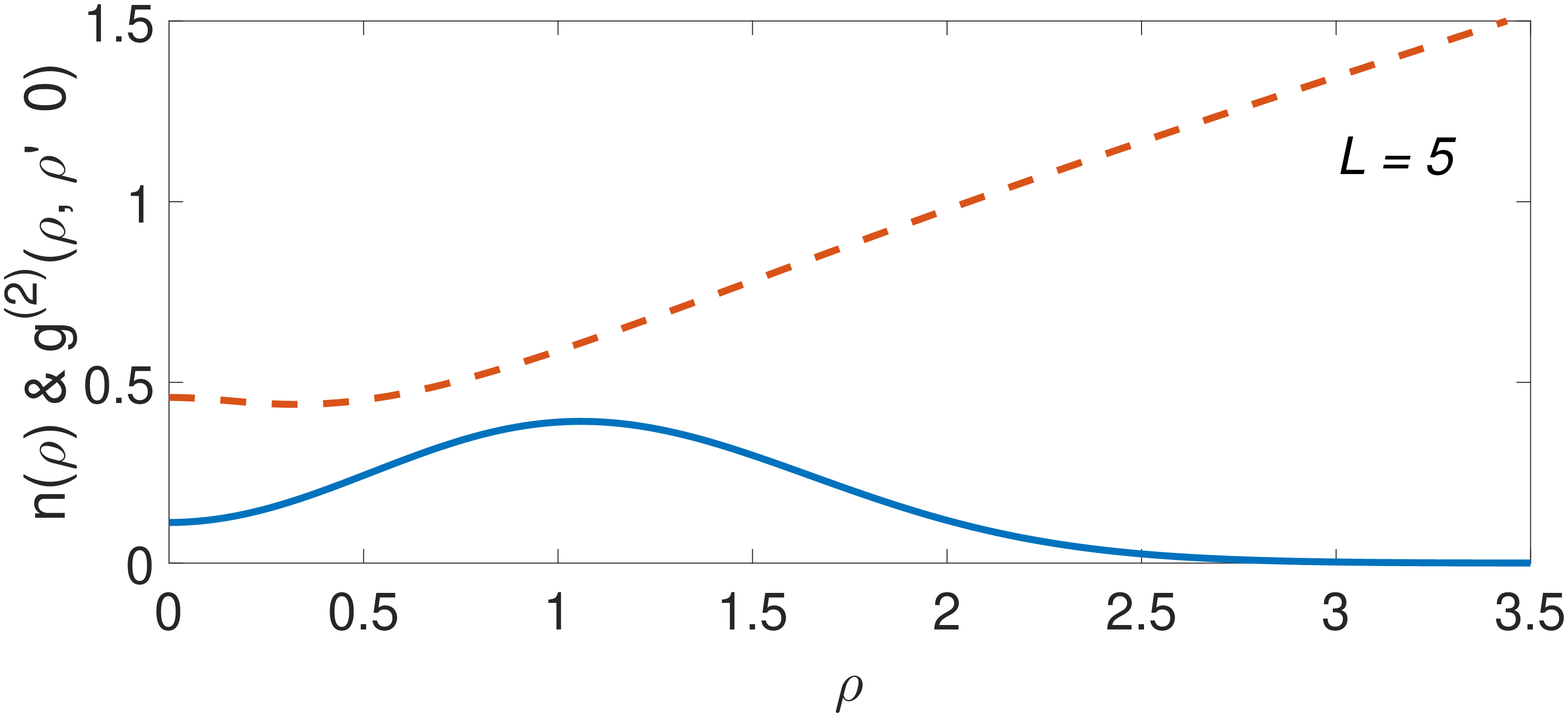}
\includegraphics[width=7cm,height=4.6cm,angle=0]{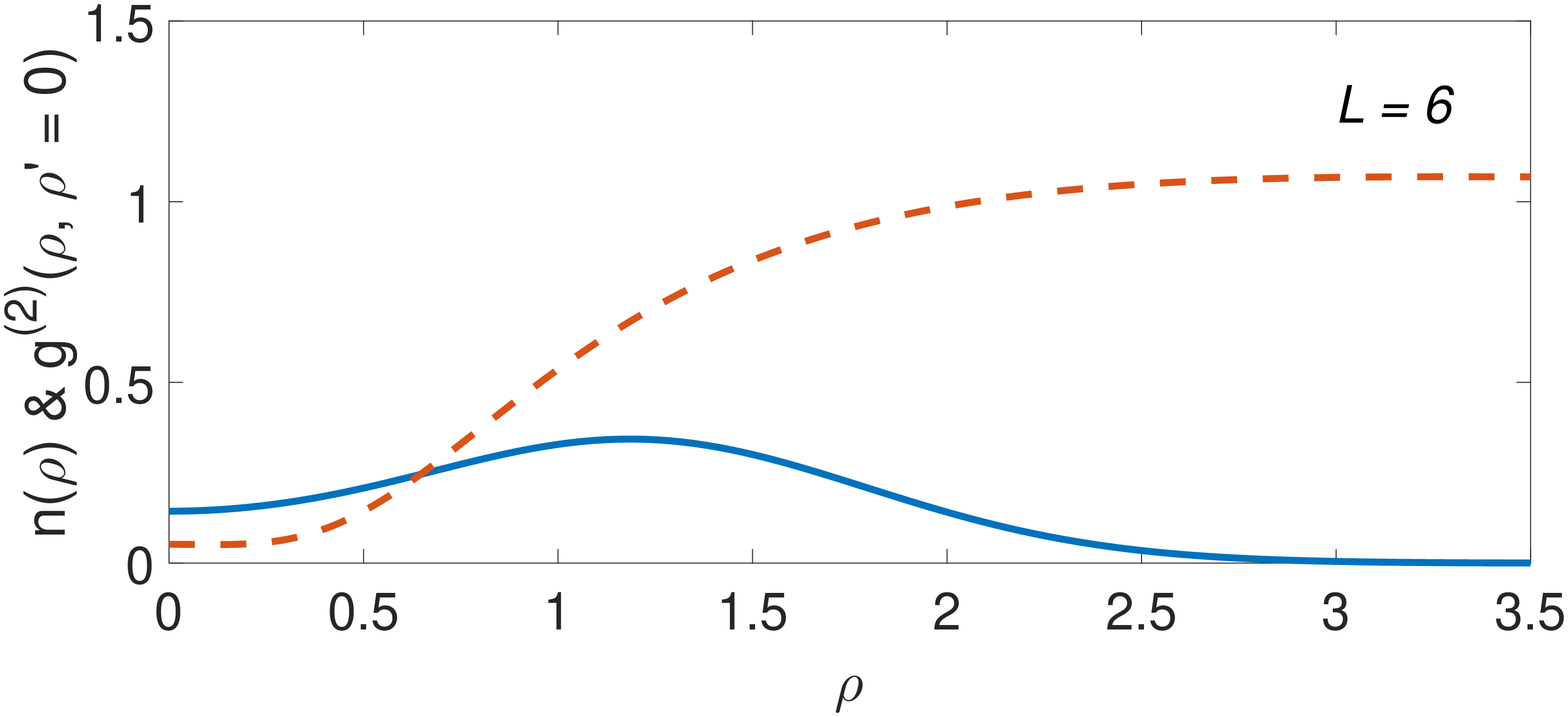}
\includegraphics[width=7cm,height=4.6cm,angle=0]{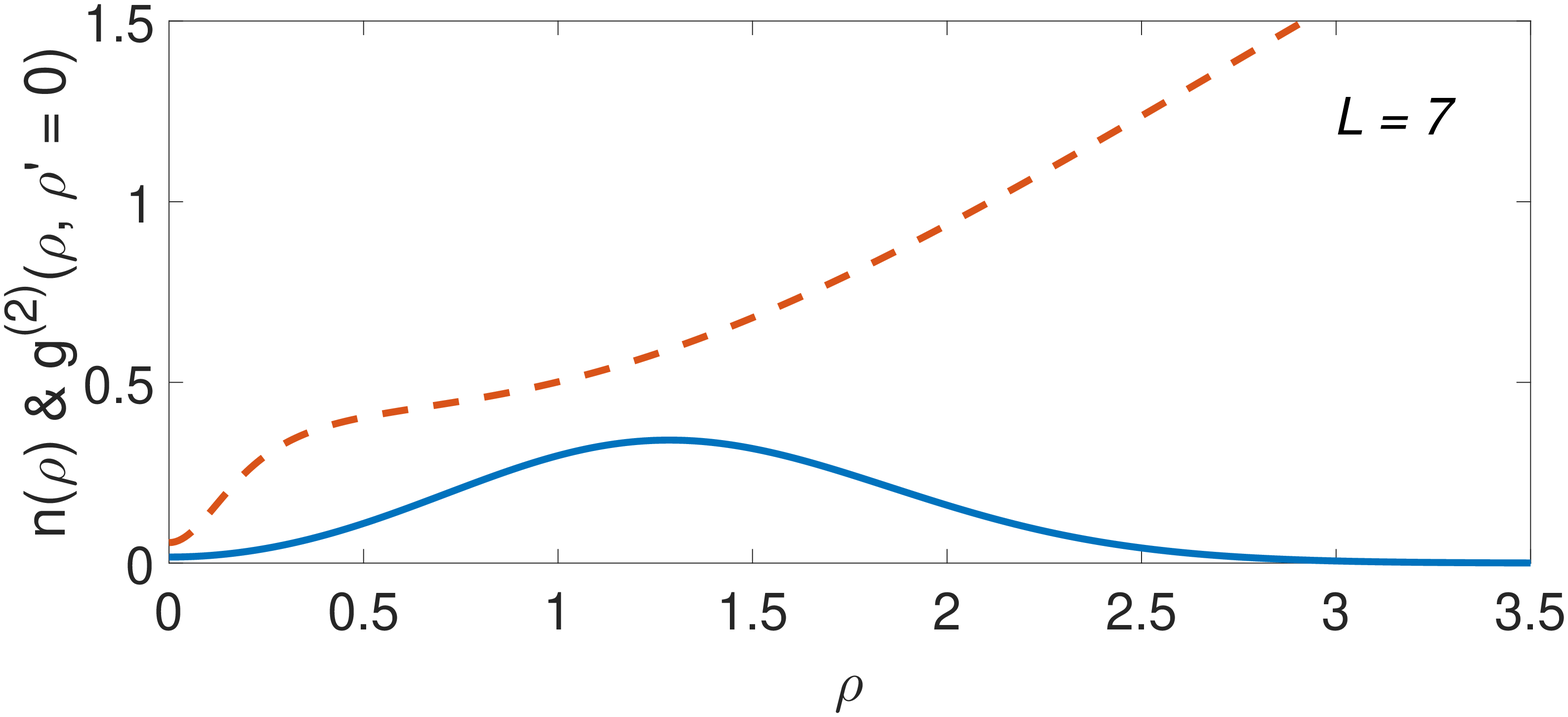}
\caption{(Color online) The density $n(\rho)$ (solid line) in units of $a_0^{-2}$, and the pair-correlation function 
$g^{(2)}(\rho, \rho' = 0)$ (dashed line) for the lowest-energy eigenstate of the Hamiltonian, for $N = 4$ atoms, and 
$4 \le L \le 7$, for $g = 0.1$ and $\lambda = 0.1$. Also $\rho$ is measured in units of $a_0$.}
\end{figure}
\begin{figure}[h]
\includegraphics[width=7cm,height=4.6cm,angle=0]{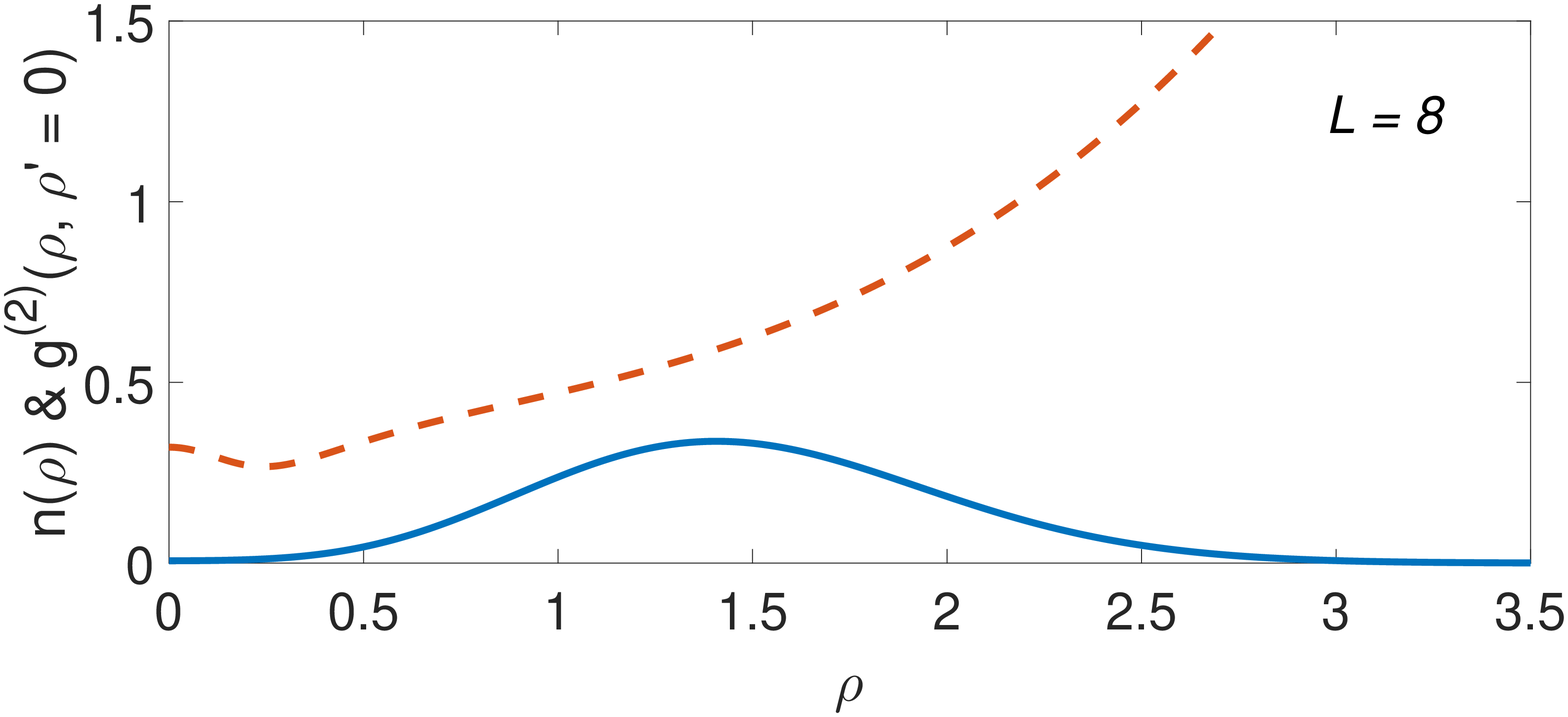}
\includegraphics[width=7cm,height=4.6cm,angle=0]{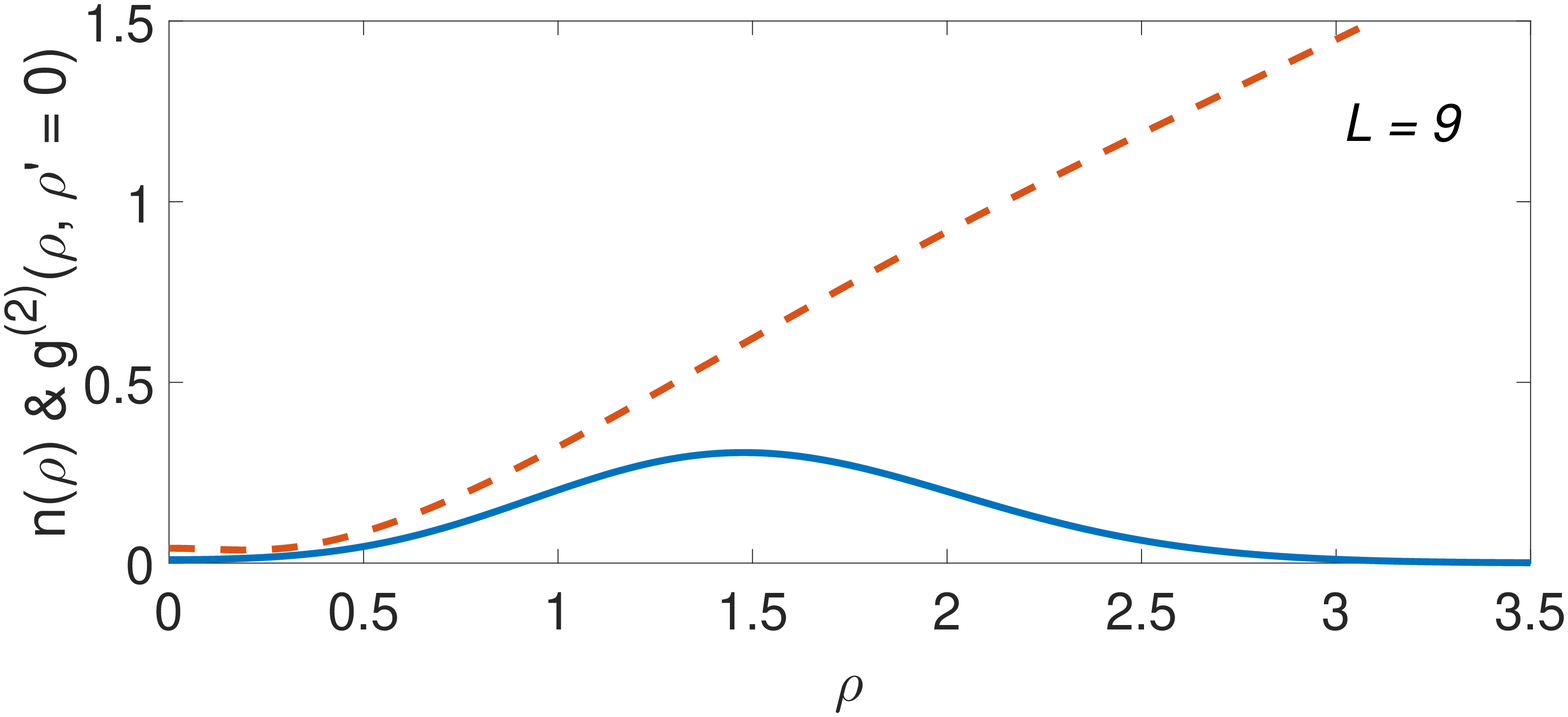}
\includegraphics[width=7cm,height=4.6cm,angle=0]{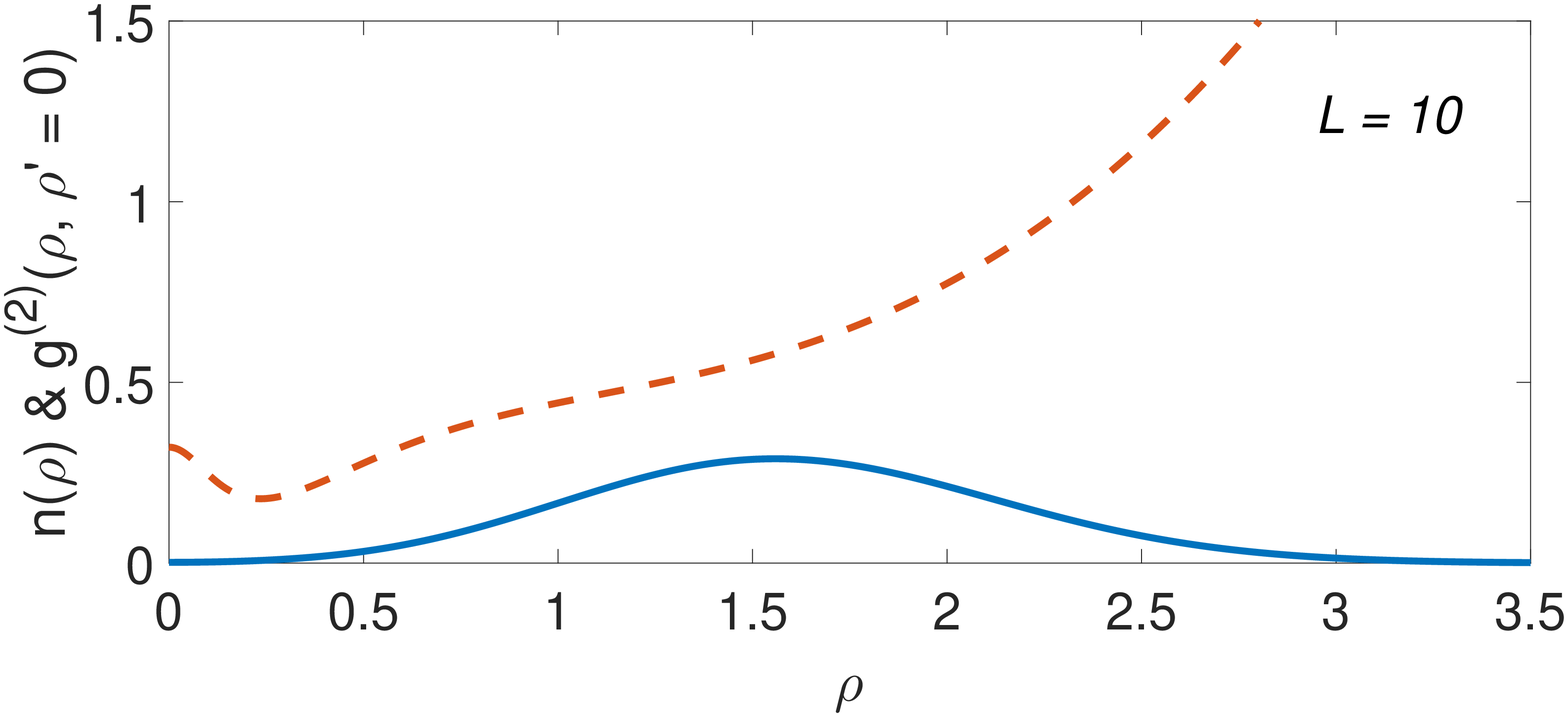}
\includegraphics[width=7cm,height=4.6cm,angle=0]{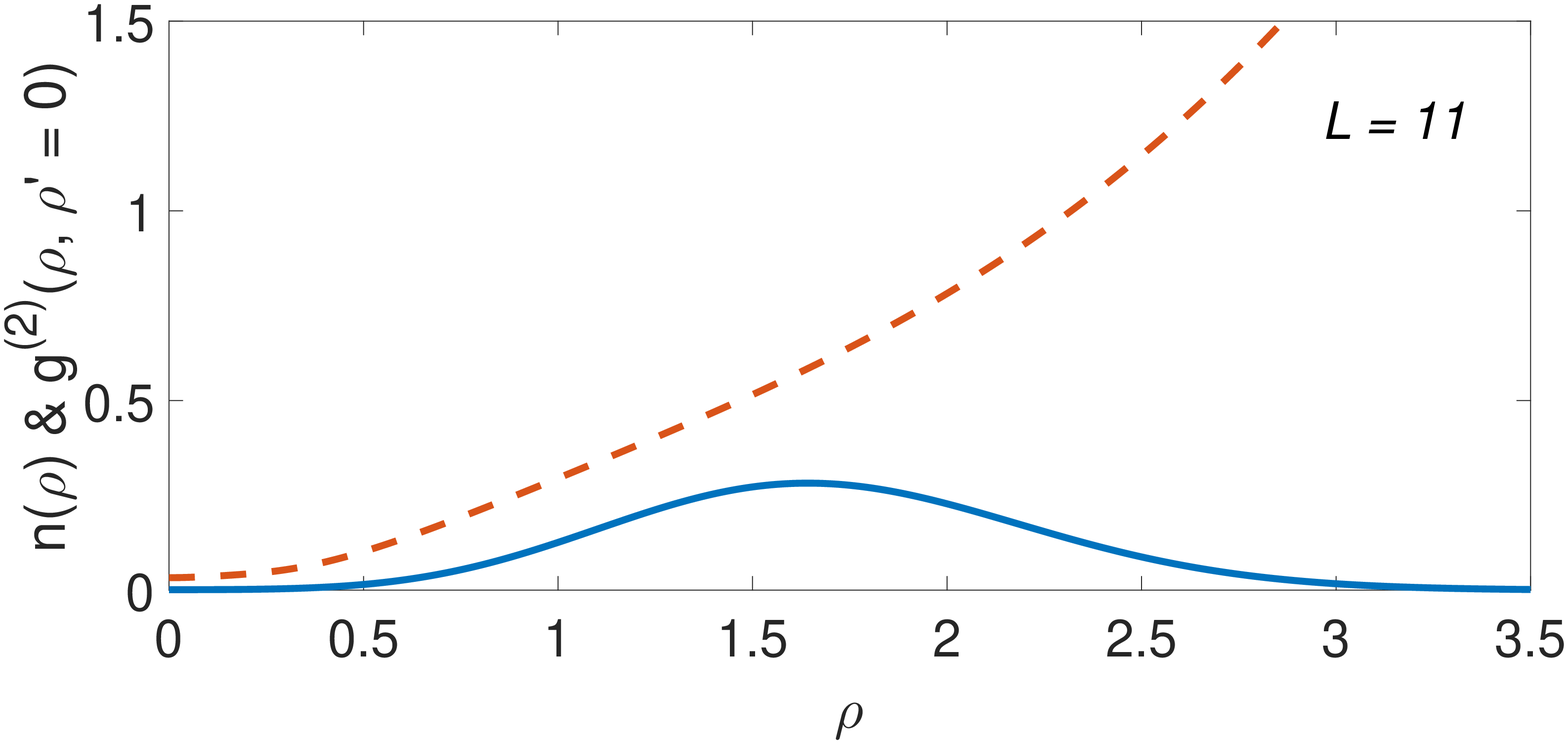}
\caption{(Color online) The density $n(\rho)$ (solid line) in units of $a_0^{-2}$, and the pair-correlation function 
$g^{(2)}(\rho, \rho' = 0)$ (dashed line) for the lowest-energy eigenstate of the Hamiltonian, for $N = 4$ atoms, and 
$8 \le L \le 11$, for $g = 0.1$ and $\lambda = 0.1$. Also $\rho$ is measured in units of $a_0$.}
\end{figure}
\begin{figure}[h]
\includegraphics[width=7cm,height=4.6cm,angle=0]{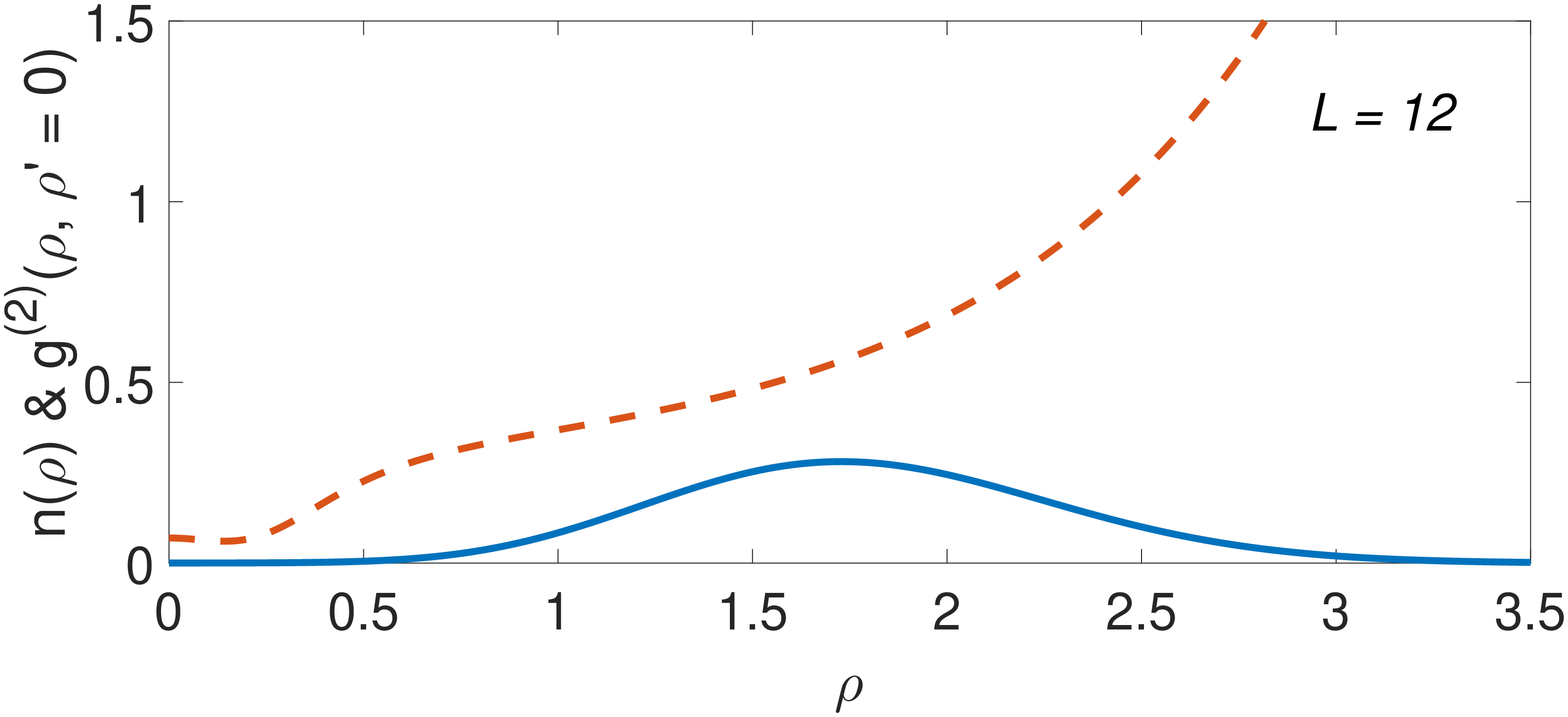}
\caption{(Color online) The density $n(\rho)$ (solid line) in units of $a_0^{-2}$, and the pair-correlation function 
$g^{(2)}(\rho, \rho' = 0)$ (dashed line) for the lowest-energy eigenstate of the Hamiltonian, for $N = 4$ atoms, and 
$L = 12$, for $g = 0.1$ and $\lambda = 0.1$. Also $\rho$ is measured in units of $a_0$.}
\end{figure}

\subsection{Anharmonic potential}

In Figs.\,13, 14, 15 and 16 we have repeated the same calculation as in Figs.\,9, 10, 11, and 12, in an anharmonic
potential, with $\lambda = 0.1$. 

While for $L$ up to 4 both the density, as well as the pair-correlation function are roughly the same, for 
$L \ge 5$ the density starts to deviate in the two cases, of harmonic and anharmonic confinement. The major
difference in the density is that while for the harmonic case it becomes more and more uniform as $L$ increases, 
in the anharmonic it is peaked around some value of $\rho$. Especially for the value of $L/N = 2$ ($L = 8$) and 
$L/N = 12$ ($L=12$) the density is rather close to the one expected for the giant-vortex states $\psi_2$ and
$\psi_3$, which have a maximum density at $\rho = \sqrt 2$ and $\sqrt 3$, respectively. 

Furthermore, for $L \ge 6$ also the pair-correlation function behaves in a different way in the two cases. 
What is even more important is that in the case of anharmonic confinement, it never develops a node, even for 
the Laughlin state (for $L=12$). This result is due to the fact that in the presence of the anharmonic potential 
the Laughlin state is highly fragile towards an uncorrelated, mean-field, giant vortex state and as a result the 
mean-field approximation is still valid \cite{frag}.

\section{Overlaps and density matrix}

Another calculation which demonstrates the transition from the mean-field regime, to the correlated regime, 
involves an overlap. More specifically we considered $N = 5$ atoms, with $g = 0.1$ and evaluated the lowest-energy 
many-body eigenstate $|\Psi \rangle$ for $0 \le L \le 20$, in a purely harmonic potential and the corresponding 
many-body state $|\Psi \rangle_{\rm anh}$ in a weakly-anharmonic potential, with $\lambda = 0.1$. 

The overlap $|\langle \Psi | \Psi_{\rm anh} \rangle|$ is shown in Fig.\,17, as function of the angular momentum
$L$. One sees that there is a general tendency of this overlap to decrease with increasing $L$. This is a direct
consequence of the transition of the system from the mean-field regime to the correlated regime. While in the 
mean-field regime the many-body state is less sensitive to the anharmonicity, as $L$ increases, the anharmonic 
part of the potential affects more and more the many-body state. 

Another interesting observation is that the overlap has local maxima for values of $L$ which are integer multiples 
of $N$, i.e., for $L = 0, 5, 10, 15$, and 20. This is due to the fact that $|\Psi_{\rm anh} \rangle$ is ``closer" 
to the mean-field giant-vortex states for these values of $L$. 

As a final result we show in Figs.\,18 and 19 the eigenvalues of the density matrix that correspond to the 
lowest-energy eigenstates of the many-body Hamiltonian for three values of the angular momentum. The eigenvalues 
of the density matrix coincide with the occupancies of the single-particle states, since the density matrix is 
diagonal (due to the axial symmetry of the problem). 

In Fig.\,18 we have considered $N=6$ atoms in a purely harmonic potential, with $g = 0.1$. For $L = 15$, i.e., 
$L/N = 2.5$, according to the mean-field approximation the state has a three-fold symmetry, where only the 
single-particle states with $m = 0, 3, 6, \dots$ are macroscopically occupied. Indeed, the states with $m = 0$ 
and $m = 3$ do have the maximum occupancy. The $m=6$ state has an occupancy which is comparable with that of
states with other values of $m$. The occupancy of these states is due to finite-$N$ corrections.

According to the mean-field approximation the pattern mentioned above (i.e., with the lowest-energy state having
some discrete rotational symmetry) continues all the way up to $L = N (N-1)$. Still, in the present 
case, where for $L = 30$ the system is in the Laughlin state, this is not the case. For example, for $L=30$ the 
occupancy of all the states with $0 \le m \le 10$ is roughly the same. For $L = 29$ a precursor of this effect 
is also seen. This effect is a direct consequence of the transition of the system from the mean-field to the 
correlated regime.

Finally, Fig.\,19 shows the same result in an anharmonic potential, with $\lambda = 0.1$. In this case
the situation changes drastically, where the same behaviour is seen for all the three values of $L$ that we have 
considered. Due to the anharmonic potential (even if it is ``weak"), there is one single-particle state $\psi_{m_0}$ 
with a large occupancy, while at least $\psi_{m_0 \pm 1}$ also have a non-negligible occupancy. Again, we see that 
the effect of the anharmonic potential is rather drastic, irrespective of the value of $L$.

\begin{figure}[h]
\includegraphics[width=8cm,height=5cm,angle=0]{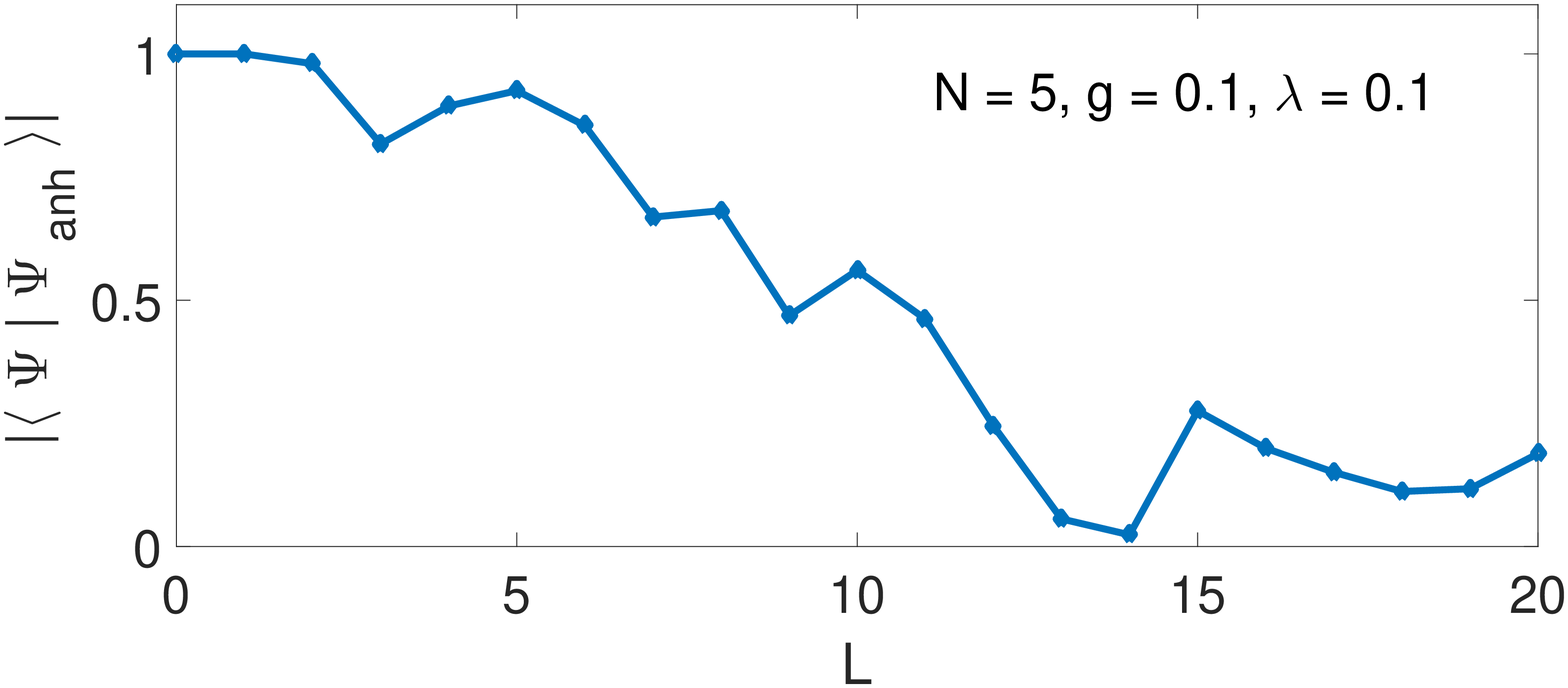}
\caption{(Color online) The overlap between the lowest-energy eigenstate $|\Psi_{\rm anh} \rangle$ of the Hamiltonian 
in a weakly anharmonic potential, with $\lambda = 0.1$, and the lowest-energy eigenstate in a purely harmonic potential
$|\Psi \rangle$, for $N = 5$ atoms, $0 \le L \le 20$, and $g = 0.1$. Here $L$ is measured in units of $\hbar$.}
\end{figure}

\begin{figure}[h]
\includegraphics[width=8cm,height=5cm,angle=0]{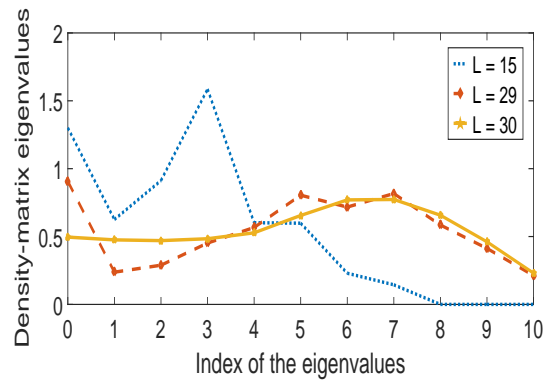}
\caption{(Color online) The eigenvalues of the density matrix in a purely harmonic potential, for $N = 6$ atoms, 
$g = 0.1$, and $L = 15$ (dotted line), 29 (dashed line), and 30 (solid line).}
\end{figure}

\begin{figure}[h]
\includegraphics[width=8cm,height=5cm,angle=0]{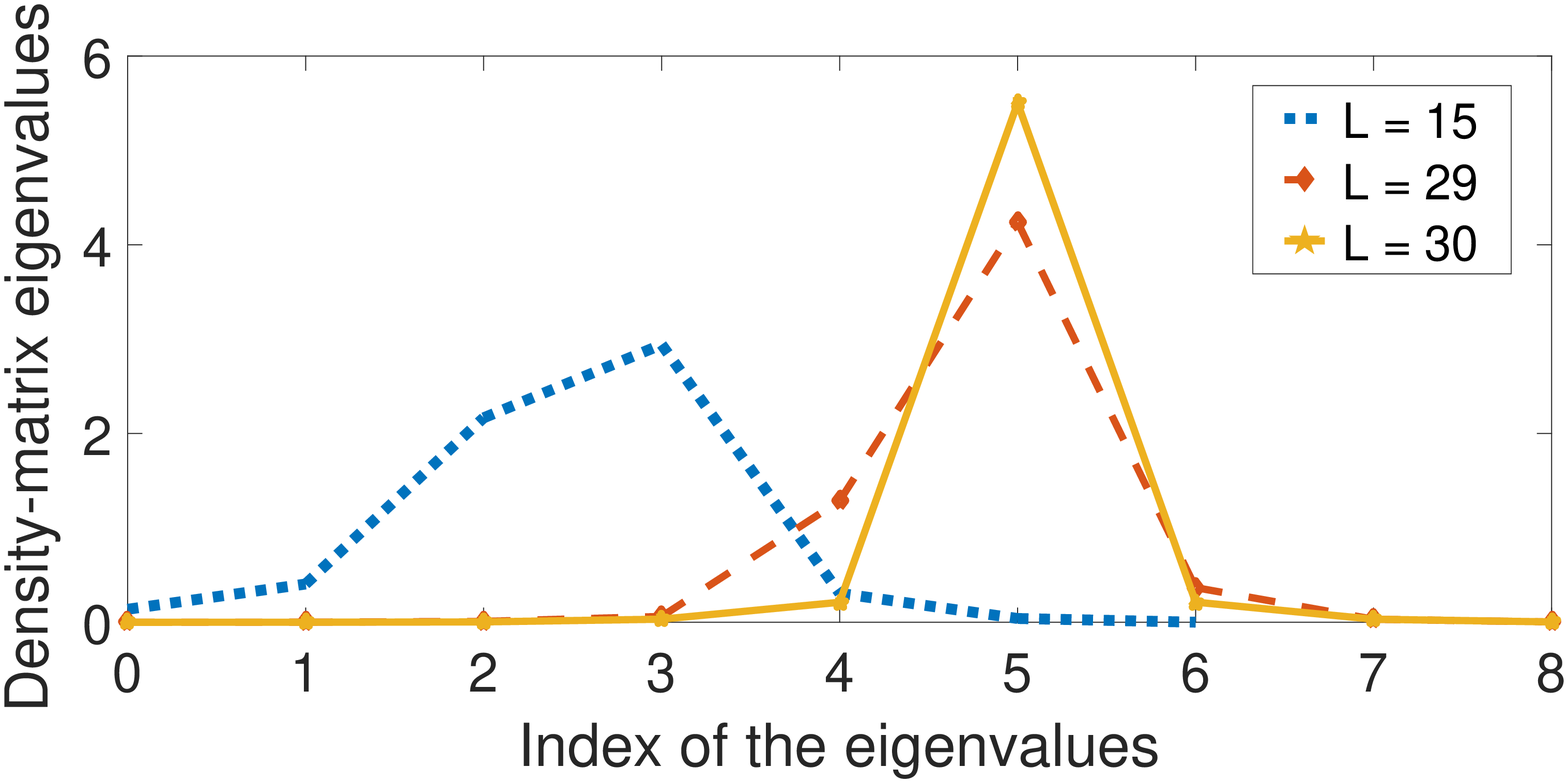}
\caption{((Color online) The eigenvalues of the density matrix in an anharmonic potential with $\lambda = 0.1$, 
for $N = 6$ atoms, $g = 0.1$, and $L = 15$ (dotted line), 29 (dashed line), and 30 (solid line).}
\end{figure}

\section{Summary and conclusions}

In the present study we have investigated numerically the rotational properties of a Bose-Einstein condensate 
which is confined either in a harmonic, or in a weakly-anharmonic trapping potential.

In the case of harmonic confinement, the system makes a transition from the mean-field regime, to a correlated 
regime, where the many-body state develops correlations that go beyond the simple product state that one assumes 
within the mean-field approximation.

Having considered a small atom number, and diagonalizing the Hamiltonian, we have evaluated the many-body state, 
which undergoes the transition mentioned above. One may identify numerous effects due to this transition, in 
connection also with the small atom number we have considered. (Clearly, for small values of $N$ the difference 
between $N$ and $N^2$ is not very significant and certainly not as significant as macroscopically-large values 
of $N$).

Firstly, the correlations introduce corrections in the energy of the gas, which are of subleading order in $N$ 
for ``slow" rotation, while they affect the energy to leading order in $N$ in the limit of ``fast rotation".
Secondly, the angular momentum $\ell(\Omega)$ of the gas for a fixed rotational frequency has a structure which 
is different than the corresponding limit of large $N$. Thirdly, the single-particle density distribution becomes
flat, while the pair-correlation function develops a node as soon as the system reaches the Laughlin state. The 
last effect is the almost equal occupancy of the single-particle states, or equivalently of the eigenvalues of the 
density matrix.

Introducing (even) a weak anharmonic potential, we observe a drastic change in the system. First of all, the energy 
is well described by the mean-field approximation, for any value of the angular momentum of the gas/angular velocity
of the trap. As a result, $\ell(\Omega)$ develops a rather regular structure, while the single-particle density 
distribution is also correspondingly simple, as well as the occupancy of the single-particle states. Correlations 
play a minor role in this case and the Laughlin state is no longer the lowest-energy state of the system in the 
limit of rapid rotation (it is a giant-vortex state, instead). Finally, the pair-correlation function no longer 
has a node, since the many-body state does not have the correlations of the Laughlin state in this case.

The derived results for small atom numbers are important for numerous reasons. First of all, there is a 
general tendency of the field of cold atoms to move towards this limit experimentally. We stress that a major 
advantage of a system with a small atom number is that one easily reaches the limit of rapid rotation, where 
the many-body state is described by the bosonic Laughlin state.

In addition, since even a weak anharmonic term in the trapping potential (which is essentially unavoidable in any
real experiment) destabilizes the correlated states, the only way for the correlated states to survive is to work 
with a small atom number \cite{frag}. As we have seen, the anharmonicity parameter affects the rotational properties 
of the system drastically. Thus, it is would be equally interesting to realize the correlated states experimentally, 
and also to investigate/confirm the effect of the anharmonic confining potential on the rotational response of
this system as the anharmonicity parameter is tuned.

\eject

\end{document}